%% file: JLT_PAS_HDD_v29_arxiv.tex
\documentclass[twocolumn,twoside,journal]{IEEEtran}
\usepackage{etex}
\usepackage{amsmath,amssymb}
\usepackage{float}
\usepackage{lipsum}
\usepackage{array}
\usepackage{cite}
\usepackage{textcomp}
\usepackage{siunitx}
\usepackage{float}
\usepackage[pdftex]{graphicx}
\usepackage{siunitx}

\usepackage{xcolor}
\usepackage{colortbl}
\usepackage{tikz}
\usetikzlibrary{decorations}
\usetikzlibrary{calc} 
\tikzset{>=latex}
\usetikzlibrary{arrows,shapes}
\usetikzlibrary{arrows}
\usetikzlibrary{plotmarks}
\usepackage{ctable}
\usepackage{pgfplots}
\usepackage[english]{babel}
\usepackage{color}
\usepackage[utf8]{inputenc}
\usepackage[T1]{fontenc}
\usetikzlibrary{decorations.pathreplacing,shapes.misc}
\usetikzlibrary{patterns}
\usepackage{tcolorbox}
\usepackage{balance}

\usepackage{mathtools}

\mathtoolsset{showonlyrefs}

\newcommand{\SNR}{\mathsf{SNR}}
\newcommand{\lab}{\mathsf{L}}
\newcommand{\inalpha}{\mathcal{X}}
\newcommand{\pyx}{p_{Y|X}}
\newcommand{\rmT}{\mathsf{T}}
\newcommand{\Bmat}{\boldsymbol{B}}

\newcommand{\x}{\boldsymbol{x}}

\renewcommand{\a}{\boldsymbol{a}}
\renewcommand{\b}{\boldsymbol{b}}
\renewcommand{\u}{\boldsymbol{u}}
\newcommand{\s}{\boldsymbol{s}}
\renewcommand{\t}{\boldsymbol{t}}
\newcommand{\p}{\boldsymbol{p}}
\newcommand{\ua}{\boldsymbol{u}^{\mathsf{a}}}
\newcommand{\us}{\boldsymbol{u}^{\mathsf{s}}}
\newcommand{\tus}{\tilde{\boldsymbol{u}}^{\mathsf{s}}}

\newcommand{\N}{\mathbb{N}}
\newcommand{\Np}{\mathbb{N}_0}

\newcommand{\rmR}{\mathsf{R}}

\renewcommand{\P}{\mathsf{P}}
\newcommand{\mI}{\mathsf{I}}
\newcommand{\ent}{\mathsf{H}}

\newcommand{\Pe}{P_\mathsf{e}}

\newcommand{\rateHDD}{\mathsf{R}_{\mathsf{HDD}}}

\newcommand{\Rs}{R_{\mathsf{s}}}
\newcommand{\nc}{n_{\mathsf{c}}}

\newcommand{\kc}{k_{\mathsf{c}}}

\makeatletter
\protected\def\vvv#1{\leavevmode\bgroup\vbox\bgroup\xvvv#1\relax}

\def\xvvv{\afterassignment\xxvvv\let\tmp= }

\def\xxvvv{%
	\ifx\tmp\@sptoken\egroup\ \vbox\bgroup\let\next\xvvv
	\else\ifx\tmp\relax\egroup\egroup\let\next\relax
	\else
	\hbox to 1.1em{\hfill\tmp\hfill}% centred
	\let\next\xvvv\fi\fi
	\next}

\makeatother
 \makeatletter
 \newcommand*{\@rowstyle}{}
 \newcommand*{\rowstyle}[1]{% sets the style of the next row
 	\gdef\@rowstyle{#1}%
 	\@rowstyle\ignorespaces%
 }
 \newcolumntype{=}{% resets the row style
 	>{\gdef\@rowstyle{}}%
 }
 \newcolumntype{+}{% adds the current row style to the next column
 	>{\@rowstyle}%
 }

\begin{document}
%\title{Probabilistically-Shaped Coded Modulation for Spectrally-Efficient Fiber-Optic Communications with Hard Decision Decoding}
\title{Probabilistic Amplitude Shaping with Hard Decision Decoding and Staircase Codes}


\author{Alireza~Sheikh \IEEEmembership{Student Member, IEEE}, Alexandre~Graell~i~Amat, \IEEEmembership{Senior Member, IEEE}, \\ Gianluigi~Liva, \IEEEmembership{Senior~Member, IEEE}, and Fabian Steiner, \IEEEmembership{Student~Member, IEEE} 
 %          \thanks{This work was presented in part at the OSA Signal Processing in Photonic Communications, Boston, MA, June-July 2015, so the current paper is the extension of \cite{Sheikh:15}.}
 \thanks{Part of this paper will be presented at the European Conference on Optical Communications (ECOC), Gothenburg, Sweden, 2017 \cite{She17c}.}
        \thanks{This work was financially supported by the Knut and Alice Wallenberg Foundation and by the Swedish Research Council under grant 2016-04253.}
            \thanks{A. Sheikh and A. Graell i Amat are with the Department of Electrical Engineering, Chalmers University of Technology, SE-41296 Gothenburg, Sweden (email: \{asheikh,alexandre.graell\}@chalmers.se).}
         \thanks{G. Liva is with the Institute of Communications and
         		Navigation of the German Aerospace Center (DLR), M\"unchner Strasse 20, 82234 We{\ss}ling, Germany
         		(email: gianluigi.liva@dlr.de).}
         \thanks{F. Steiner is with the Institute of Communication Engineering of the Technical University of Munich (TUM), Theressienstrasse 90,
         	80333 Munich, Germany
         	(email: fabian.steiner@tum.de).}
         
        }

\maketitle

\begin{abstract}
 
We consider probabilistic amplitude shaping (PAS) as a means of increasing the spectral efficiency of fiber-optic communication systems. In contrast to previous works in the literature, we consider probabilistic shaping with hard decision decoding (HDD). In particular, we apply the PAS recently introduced by B\"ocherer \emph{et al.} to a coded modulation (CM) scheme with bit-wise HDD that uses a staircase code as the forward error correction code. We show that the CM scheme with PAS and staircase codes yields significant gains in spectral efficiency with respect to the baseline scheme using a staircase code and a standard constellation with uniformly distributed signal points. Using a single staircase code, the proposed scheme achieves performance within $0.57$--$1.44$ dB of the corresponding achievable information rate for a wide range of spectral efficiencies.

\end{abstract}


\thispagestyle{empty} \setcounter{page}{0}

\section{Introduction}\label{int}

\IEEEPARstart{T}{o meet} the ever increasing data rate demands, next-generation fiber-optic communication systems need to use the available spectrum more efficiently. Therefore, there is currently a great focus in the research community in increasing the spectral efficiency of these systems. In this regard, forward error correction (FEC) in combination with higher order modulation, a scheme commonly referred to as coded modulation (CM), has become a key part of fiber-optic systems.

Traditionally used signal constellations, such as amplitude shift keying (ASK) and quadrature amplitude modulation (QAM), are characterized by equidistant signal points and  uniform signaling, i.e., each signal point is transmitted with the same probability. Unfortunately, such constellations result in a gap to the Shannon limit ($1.53$ dB for an additive white Gaussian noise (AWGN) channel in the high signal-to-noise ratio (SNR) regime). To close this gap and to increase the spectral efficiency, signal shaping may be applied. There are two main classes of signal shaping, geometric shaping \cite{Barsoum_2007,polar_modulation,Djordjevic_2010,Liu_2014,Geller_2016}, and probabilistic shaping \cite{Calderbank_1990,Forney_1992,georg_tcom}. In geometric shaping, the constellation points are arranged in the complex plane in a nonequidistant manner to mimic the capacity achieving distribution. Probabilistic shaping, on the other hand, starts with a constellation with equidistant signal points, e.g., ASK or QAM, and assigns different probabilities to different constellation points.

Both geometric and probabilistic shaping have been considered for fiber-optic communications as a means to increase the spectral efficiency, showing significant gains with respect to conventional constellations \cite{Geller_2016,Smith_2012,Buchali1_2015,Pan,Fehenberger}. A significant advantage of probabilistic shaping is that it builds up on off-the-shelf constellations, hence incurring no additional complexity in system design and implementation compared to geometric shaping. In \cite{georg_tcom}, a new CM scheme using probabilistic shaping was proposed. The proposed scheme, dubbed probabilistic amplitude shaping (PAS), was shown in \cite{georg_tcom} to achieve performance within $1.1$ dB from the capacity of the AWGN channel for a wide range of spectral efficiencies using off-the-shelf low-density parity-check (LDPC) codes and soft decision decoding (SDD). More recently, this scheme has been considered for fiber-optic communications in \cite{Buchali_2016,Gha17}.  

All these previous works consider shaping in combination with SDD. However, while FEC with SDD yields very large net coding gains, it entails a high decoding complexity, which translates into a large chip area and power consumption \cite{pillai_2014_jlt}. To reduce the decoding complexity, hard decision decoding (HDD) is an appealing alternative. Hard decision decoders consume significantly less power than soft-decision decoders \cite{pillai_2014_jlt}. Despite the rise of FEC with SDD, such as LDPC codes and spatially-coupled LDPC codes \cite{Hager15}, the interest in FEC schemes with HDD has experienced a revived attention in the fiber-optic communications research community in the past few years, thanks to the appearance of very powerful FEC-HDD schemes. Constructions such as staircase codes \cite{staircase_frank,christian1,She17}, braided codes \cite{h_braided}, and other generalized product codes \cite{Hag17,hager_coded_mod} yield very large net coding gains yet with much lower decoding complexity than FEC-SDD schemes.

In this paper, we consider signal shaping as a means of increasing the spectral efficiency of the fiber-optic system without increasing the launch power. However, contrary to the previous literature on signal shaping,  which focuses on SDD, we consider signal shaping with (bit-wise) HDD. In particular, we apply the PAS scheme proposed in \cite{georg_tcom} to a CM scheme with staircase codes and HDD. Similar to \cite{georg_tcom}, using ASK modulation as the underlying signal constellation, we optimize shaping such that a given achievable information rate, with bit-wise HDD in our case, is maximized. 
This paper extends the work presented in \cite{She17c}, where the application of PAS to bit-wise HDD was originally proposed, showing remarkable coding gains. In \cite{She17c} the design was based on the maximization of the generalized mutual information. Here, we aim at optimizing the input distribution for a rate that is achievable by the characteristic decoding strategy of PAS, derived in \cite{bocherer2017}. 
We then discuss the adaptation of  PAS to the use of staircase codes and the optimization of the code parameters. Furthermore, we address the selection of the operating point for finite frame length. 
%We show that the optimization of the shaping based on maximizing the MI yields to a practical operating point for finite frame length far from the theoretical limit for HDD. Thus, we propose a slight modification to the shaping which results in a practical operating point with performance closer to the theoretical limit for HDD. 
Finally, we show through simulation results that the probabilistic shaping CM scheme with staircase codes and HDD achieves up to $2.88$~dB gain improvement with respect to the system using a staircase code and a conventional, uniform signal constellation.% over a wide range of spectral efficiencies.

The remainder of the paper is organized as follows. In Section~\ref{GMIdis}, the achievable information rate with bit-wise HDD is discussed. In Section~\ref{shaped_sec}, the PAS scheme with HDD is introduced. The CM scheme with PAS and staircase codes is further elaborated in Sections~\ref{coded_modulation} and~\ref{operating_point}. Finally, simulation results are given in Section~\ref{sec:simulation} and some conclusions are drawn in Section~\ref{conclusion}.
 
Notation: The following notation is used throughout the paper. We define the sets $\N \triangleq \{1,2,\ldots\}$ and $\Np\triangleq \{0,1,\ldots\}$. We denote by $P_X(\cdot)$ the probability mass function (pmf) and by $p_X(\cdot)$ the probability density function (pdf) of a random variable (RV) $X$. We use boldface letters to denote vectors and matrices, e.g., $\boldsymbol{x}$ and $\boldsymbol{X}$, respectively. Expectation with respect to the pmf of RV $X$ is denoted by $\mathbb{E}_X(\cdot)$. $\ent(X)$ and $\mI(X;Y)$ stand for entropy of the RV $X$ and mutual information between RVs $X$ and $Y$, respectively.

%\subsection{Notation}
%${\hat X}$ represents the output of the hard detection applied to the output of the channel when the input is $X$.

%\section{Preliminaries}
%\label{sec:preliminaries}
\section{Achievable Information Rate with Hard Decision Decoding and Bit-Wise Decoding}\label{GMIdis}

We consider a discrete-time AWGN channel\footnote{The AWGN channel is an accurate model for long-haul coherent fiber-optic communications when the fiber-optic channel is dominated by amplified spontaneous emission noise \cite{Pog12}.} with input-output relation at time instant \emph{i}
\begin{align*}
%\label{in_outAWGN}
{Y_i} = \Delta{X_i} + {Z_i} \qquad\qquad i=1,2,\ldots,n
\end{align*}
where $n$ is the number of channel uses (i.e., the block length), ${X_i}$ is the input of the channel, ${Y_i}$ is its output, $\Delta$ is a scaling constant, and $\{Z_i\}$ are independent and identically distributed (i.i.d.) Gaussian RVs with zero mean and unit variance. The scaling parameter $\Delta$ is defined to attain an average transmit power $\P$ according to 
\begin{align}
\mathbb{E}\left[(\Delta X)^{2}\right]=\P. \label{scaled_const2}
\end{align}

According to the definitions above, the SNR is given by $\SNR=\P$. We consider a block-wise transmission system where $\u$ denotes the transmitted information block and $\hat \u$ denotes the decoded information block. %The block error probability of the system is defined as $\Pe = \Pr(\u\ne\hat{\u})$.

For simplicity, for the analysis and the design of the PAS scheme in this and next section, we consider ASK modulation as the underlying modulation, i.e., the channel input alphabet is given by $\mathcal{X}\triangleq\{-2^m+1,...,-1,1,...,2^m-1\}$, where $m$ is the number of bits per symbol, and $M=2^m$ is the number of signal points. However, the PAS scheme directly extends to square QAM constellations, which can be seen as the Cartesian product of two ASK constellations (see also Section~\ref{code_operating}). In Section~\ref{sec:simulation} we give results for QAM constellations.

For a given distribution $P_X$ of the channel input, the mutual information (MI)
 between the channel input $X$ and channel output $Y$
 	\begin{align}
 	\mathsf{I}(X;Y)\triangleq\mathbb{E}\left[ \log_2 \left( \frac{\pyx(Y|X)}{\sum_{x'\in\inalpha}\pyx(Y|x')P_X(x')}\right)  \right]\label{eq:MI}
 	\end{align}
  determines the upper limit on the achievable rate. A rate $\rmR$ is achievable, i.e., the probability of error can be made arbitrarily small in the limit of infinitely large block length $n$, if $\rmR < \mathsf{I}\left( {X;Y} \right)$. 
The ultimate limit is given by the channel capacity, obtained maximizing the MI over all possible input distributions, $\mathsf{C}\triangleq\sup_{P_X} \mI(X;Y)$.

In this paper, we consider the PAS scheme of \cite{georg_tcom} with a binary code for transmission and bit-wise HDD (i.e., bit-wise Hamming metric decoding) at the receiver side.
%For a given channel, detection method, and decoding metric, it is possible to derive an achievable  rate for that specific system \GL{\cite{bocherer2017}}.
Denote by $\hat{X}\in \mathcal{X}$ the RV associated with the detector output (hard decision). An achievable rate for the PAS scheme can be computed by resorting to the approach introduced in \cite{bocherer2017}, yielding
	\begin{align}
	\rateHDD \hspace{-0.1em}= \mathop {\sup }\limits_{s > 0} \left[\ent\left(X\right)+\mathbb{E}\left[  \displaystyle \displaystyle\log_2 \frac{{q(X,\hat{X})}^s}{\displaystyle \sum_{x'\in\inalpha}{q(x',\hat{X})}^{s}} \right]\right]^{+} \label{GMI_rate}
	\end{align}
where $(a)^{+}=\text{max}(0,a)$, $s$ is the optimization parameter, and $q(X,\hat{X})$ is the (mismatched) decoding metric \cite{kaplan1993information,mismatch_lapidoth}. For HDD, the decoding metric is the bit-wise Hamming metric, which has the equivalent form
	\begin{align} \label{hamming_metric}
	q(x,\hat{x}) = \varepsilon ^{{\mathsf{d}_\mathsf{H}(\lab(x),\lab(\hat{x}))}}
	\end{align}
	where $\varepsilon$ is an arbitrary constant in $(0,1)$, $\lab(x)$ is the $m$-bit labeling associated with constellation symbol $x$, and ${{\mathsf{d}_\mathsf{H}}(\lab(x),\lab(\hat{x}))}$ is the Hamming distance between the binary labelings of $x$ and $\hat x$. Here, we consider the binary reflected Gray code (BRGC) labeling \cite{Gra53}.
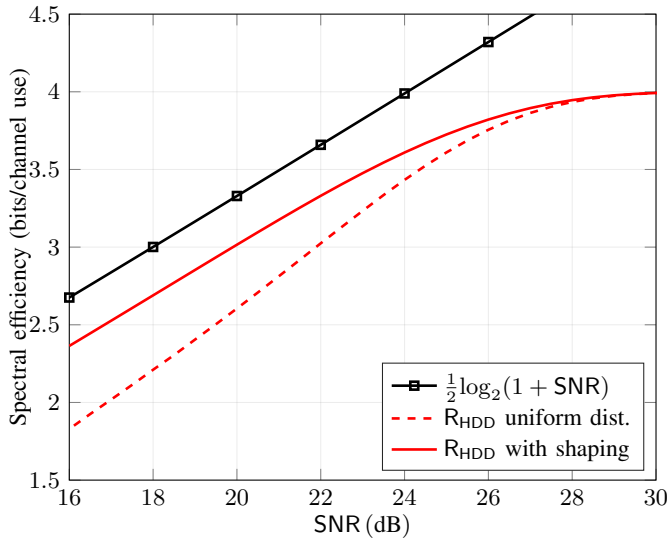
\begin{figure}[t] \centering  \newlength\figureheight
	\newlength\figurewidth
	\setlength\figureheight{6.2cm}
	\setlength\figurewidth{7.8cm} \input{MI_GMI_16ASK_without_tol_new_MAP_georgrate.tikz} 
	\vspace{-5ex}
	\caption{Achievable rates with bit-wise HDD ($\rateHDD$ given by \eqref{GMI_rate}) with and without shaping for a $16$-ASK constellation (for the uniform case, see also\cite{She17b,SheJLT}). The unconstrained-input AWGN channel capacity is also shown as reference.} \vspace{-2ex}
	\label{FF1} 
\end{figure}
By optimizing over $s$, it is possible to show that \eqref{GMI_rate} reduces to \cite[Sec.~6.4]{bocherer2017}
	\begin{align}
	\rateHDD \hspace{-0.1em}= \left[\ent\left(X\right)-m\mathsf{H}_b(p)\right]^{+}\label{GMI_rate2}
	\end{align}
	where $\mathsf{H}_b(p)=-p\log_2 p -(1-p)\log_2(1-p)$ is the binary entropy function, and $p$ is the (raw) bit error probability at the output of the hard detector (i.e., the pre-FEC bit error rate).
In Fig.~\ref{FF1}, we plot $\rateHDD$ for a $16$-ASK constellation with uniform distribution. For the sake of comparison, we also plot the capacity of the unconstrained-input AWGN channel.

\section{Probabilistically-Shaped ASK Constellation with Hard Detection}\label{shaped_sec}

In this section, we use probabilistic shaping to boost the achievable rate compared to the case where constellation points are drawn uniformly from $\mathcal{X}$. 
Similar to \cite{georg_tcom}, we consider the Maxwell-Boltzmann distribution for the channel input $X$,
 \begin{align} \label{maxwell_boltzmann}
 {P^{\lambda}_{X}}\left( x \right) = \frac{\exp\left( - \lambda x^2\right)}{\sum\limits_{\tilde{x} \in \mathcal{X}} \exp\left( - \lambda \tilde{x}^2\right) }.
 \end{align}
 For each SNR, we select $\lambda$ such that the achievable rate is maximized, i.e.,
\begin{align}
\label{MI_rate}
\lambda^{*}=\mathop {{\text{argmax}}}\limits_{\lambda} \;  \rateHDD.
\end{align}

Unlike \cite{georg_tcom}, in this paper we consider HDD at the receiver. In particular, we consider a symbol-wise maximum a-posteriori (MAP) detector that outputs
\begin{align}
\hat{x}= \mathop{\text{argmax}}\limits_{x \in \mathcal{X}} \;  p_{Y|X}(y|x)P_{X}(x)
\end{align}	
yielding the conditional pmf $P_{\hat X|X}$ to be used, jointly with $P_X$, to compute \eqref{GMI_rate}.

\begin {table}[!t]
\renewcommand{\tabcolsep}{0.15cm}
\caption {Shaping gain for different Spectral efficiencies} \label{T1}
\vspace{-0.25cm}
\begin{center}
	%\scriptsize
	\vspace{-2ex}
	\begin{center}\begin{tabular}{ccc}
			\arrayrulecolor{black}\hline
			\toprule

			Modulation & SE (bit/channel use)  & 	Shaping gain (dB) \\
            
			\midrule
            $4$-ASK & 1  & 	0.78 \\
%            $4$-ASK & 1.5  & 	0.67 \\
            $8$-ASK & 2  & 	1.56 \\
%            $8$-ASK & 2.5  & 	1 \\     
            $16$-ASK & 3  & 	1.98 \\
%            $16$-ASK & 3.5  & 	1.19 \\
            $32$-ASK & 4  & 	2.22 \\
%            $32$-ASK & 4.5  & 	1.28 \\               
            $64$-ASK & 5  & 	2.37 \\
%            $64$-ASK & 5.5  & 	1.35 \\             
   			\hline
   			\toprule                			
		\end{tabular} \end{center}
	\end{center}
	\vspace{-0.5cm}
\end{table}

\begin{figure}[t] \centering 
	\setlength\figureheight{6.2cm}
	\setlength\figurewidth{7.8cm}  \input{differentASKorder_updatenew_georgrate.tikz} 
	\vspace{-5ex}
	\caption{Achievable rates of the probabilistic amplitude shaping with bit-wise HDD. The red points show the SNR values after which a higher order modulation should be used. Below the red dots, the achievable rate curve for a given ASK constellation virtually overlaps with that of the lower-order constellation.} \vspace{-2ex}
	\label{FF2} 
\end{figure}
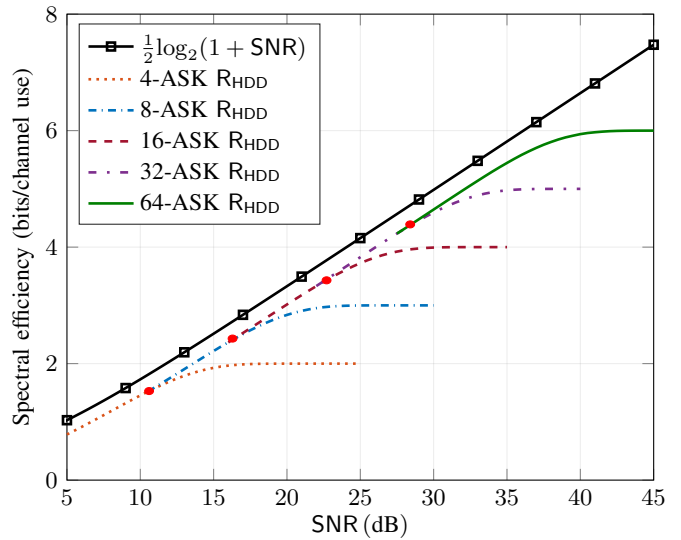

\begin{figure*}[!t]
	\centering
	\vspace{-0.4cm}
	\scalebox{0.8}{
		\begin{tikzpicture}[>=latex']
		\tikzset{Source/.style={rectangle, draw, thick, minimum width=1.6cm, minimum height=1cm, rounded corners=2mm}}
		\tikzset{Sourcehatch/.style={rectangle, draw, thick, minimum width=1.6cm, minimum height=1cm, rounded corners=2mm}}		
		\tikzset{Sourcerealimag/.style={rectangle, draw, thick, minimum width=1cm, minimum height=0.3cm, rounded corners=2mm}}		
		\tikzset{Destination/.style={rectangle, draw, thick, minimum width=1.6cm, minimum height=1cm, rounded corners=2mm}}
		\tikzset{Noise/.style={circle, draw, thick, scale=0.5, minimum size=0.1mm}}
		\tikzset{descr/.style={fill=white}}
		\tikzset{Source1b/.style={rectangle, draw=black, thick, minimum width=0.5cm, minimum height=1.2cm, rounded corners=0.5mm}}
		\tikzset{Source2/.style={rectangle, draw, thick, minimum width=0.8cm, minimum height=1.2cm, rounded corners=0.5mm}}
		\tikzset{Sourcequad/.style={rectangle, draw, thick, dashed, red, minimum width=1.6cm, minimum height=1cm, rounded corners=2mm}}									
		% Node generation
		%Uniform and Shaped blocks
		\node[Source,fill=gray!20,thick,align=center] (C1) at (-22.2,3) {Source\\(Uniform)};
		\node[Sourcehatch,fill=gray!20,thick] (C2) at (-19.7,3) {\small CCDM};
		\node[] (B2) at (-20.2,,-5.8) {\large $a_1,...,a_n$};			
		%Mapping
		\node[Source2,fill=gray!20,thick] (C3) at (-17.18,1.75) {\large $\Phi_{\mathrm{ab}}$};
		%		\node[Source,fill=red!30!white] (C3) at (-15.5,1.7) {Mapping};
		\node[Source,fill=gray!20,thick] (C4) at (-12.6,1.7) {\small Encoder};
%		\node[Sourcerealimag,fill=red!30!white] (I) at (-12.6,3.3) {$j\mathcal{I(\cdot)}$};	
%		\node[Sourcerealimag,fill=red!30!white] (R) at (-12.6,2.75) {$\mathcal{R(\cdot)}$};				
%		\node[Source,fill=red!30!white] (C5) at (-7.6,1.7) {De-mapping};
		\node[Source2,fill=gray!20,thick] (C5) at (-8.8,1.7) {\large $\Phi_{\mathrm{bs}}$};
%		\node[Noise] (C88) at (-6.7,3) {\huge $\times$};
		\node[Noise] (cc9) at (-4.7,3) {\huge $\times$};
		\node[Noise] (imag) at (-6.8,3) {\huge $\times$};
%		\node[Noise] (imag1) at (-4.87,3) {\huge $+$};				
		\node[] (cc91) at (-5,3) {};
		%		\node[Source] (C6) at (-2,1.7) {Encoder};
		\node[] (C7) at (-13.95,3) {};  %connection to real/imag
		\node[] (C77) at (-14.08,2.57) {};
		\node[] (C78) at (-14.08,3.45) {};				
		%		\node[] (B1) at (-20.6,2.1) {\Large $u_1,\ldots,u_{k-\gamma n}$};
		\node[] (B1) at (-20.1,2.1) {\large $\ua=(u_1,\ldots,u_{k-\gamma n})$};
		\node[] (D1) at (-22.2,0.6) {};	
		\node[] (D2) at (-22.31,0.725) {};	
		\node[] (D3) at (-14.6,0.725) {};	
		
		%		\node[] (B1) at (-20.6,,0.09) {\Large $u_1,...,u_{\gamma n}$};	
		\node[] (B1) at (-20.2,1.1) {\large $\us=(u_1,...,u_{\gamma n})$};	
		\node[] (B2) at (-15.3,,-2.5) {\large $\b_1,...,\b_n$};	
		\node[] (B3) at (-11.5,,-2.5) {\large $\p_1,...,\p_{n(1-\gamma)}$};	    	
		\node[] (B4) at (-8.5,,-2.4) {\large $s_1,...,s_{n}$};	
%		\node[] (B4p) at (-7.6,,-1.4) {\large $s_{n+1},...,s_{2n}$};			
		\node[] (B5) at (-8,,-5.9) {\large $x_1,...,x_n$};		        	
		%receiver
		\node[] (C8) at (-18.2,3.12) {};	    % connection to mapping	
		\node[] (D4) at (-18.2,1.56) {};	
		\node[] (D5) at (-18.33,1.68) {};
		\node[] (D6) at (-17.45,1.68) {};

		\node[] (R1i) at (-12.3,3.5) {};
		\node[] (R1o) at (-5.95,3.5) {};

		\node[Source1b,fill=gray!20,thick] (M1) at (-15.5,1.7) {};
		\node[black] at (-15.5,1.78) {\small \vvv{\sffamily p/s}};	
		
		\draw [->] (-16.75,2.25) --  node [above] {\small $b^{i}_1$} (-15.75,2.25);
		\draw [->] (-16.75,2.25) --  node [below] {\small $\vdots$} (-15.75,2.25);		
		\draw [->] (-16.75,1.2) --  node [above] {\small $b^{i}_{m-1}$} (-15.75,1.2);
		%        \node[blue] (-16.4,1.49) {$\vdots$}         
				
	     \node[Sourcequad,fill=red!20,thick,align=center] (beFOC) at (-3,3) {Quadrature\\Multiplexer};	
		\draw [->] (C1) -- (C2);
		\draw [->] (C2) -- (imag);
		\draw [->] (imag) -- (cc9);				
		\draw [->] (cc9) -- (beFOC);				
		\draw [-] (C1) -- (D1);		
		\draw [-] (D2) -- (D3);	
		\draw [-] (C8) -- (D4);			
		\draw [->] (D5) -- (D6);	
		\draw [->] (M1) -- (C4);
%		\draw [-] (C77) -- (C78);
		
%		\draw [->] (R1i) -- (R1o);	
		%		\draw [->] (-5.7,3.5) -- (-5.7,3.25);	
%		\draw [-] (-12.12,2.8) -- (-8,2.8);		
%		\draw [-] (-8,2.8) -- (-8,3);
%		\draw [->] (-8,3) -- (-6.95,3);		
		
%		\draw [->] (-14.07,2.7) -- (-13.1,2.7);
%		\draw [->] (-14.07,3.32) -- (-13.1,3.32);							
		%		\draw [->] (C3) -- (C4);			
		\draw [->] (-12.54,0.73) -- (-12.54,1.2);
		\draw [-] (-14.74,0.73) -- (-8.84,0.73);
		\draw [->] (-8.85,0.73) -- (-8.85,1.1);				
		\draw [->] (C4) -- (C5);	
%		\draw [-] (-8.35,2.1) -- (-6.7,2.1);
%		\draw [->] (-6.7,2.1) -- (-6.7,2.7);	
		\draw [-] (C5) -- (-6.75,1.7);
		\draw [->] (-6.75,1.7) -- (-6.75,2.7);		
%		\draw [->] (-6.7,2.1) -- (-6.7,2.7);			
%		\draw [->] (C88) -- (cc91);	
%		\draw [->] (imag1) -- (cc9);	
%		\draw [-] (imag) -- (-4.85,3.5);	
%		\draw [->] (-4.85,3.5) -- (-4.85,3.25);					
		\draw [->] (-4.7,1.7) -- (-4.7,2.75);	
		\node[] (D10) at (-4.65,1.5) {$\Delta$};
		\node[cloud, cloud puffs=14.7, cloud ignores aspect, minimum width=1.3cm, minimum height=1cm, align=center, draw,fill=gray!20,thick] (cloud) at (-0.9,3) {FOC};		
		\draw [->] (beFOC) -- (cloud);		
																				
		\end{tikzpicture}
	}
	\caption{Block diagram of the probabilistically amplitude shaped CM scheme. 
		%$\mathcal{R(\cdot)}$ takes the real part while $\mathcal{I(\cdot)}$ takes the imaginary part.
		}
	\label{figsystemmodel}
	\vspace{-0.4ex}
\end{figure*}
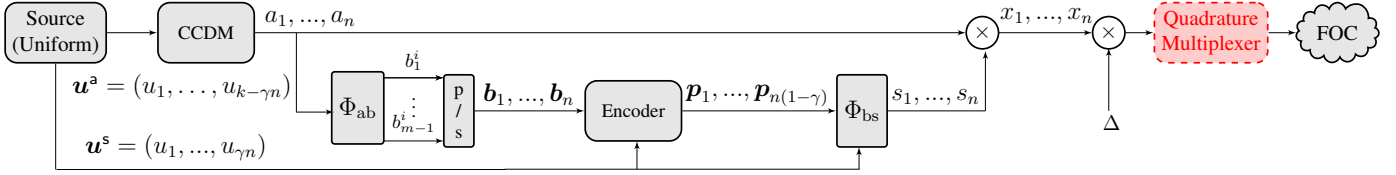

In Fig.~\ref{FF1}, we depict the achievable rate in \eqref{GMI_rate} for the probabilistically-shaped scheme according to \eqref{maxwell_boltzmann}--\eqref{MI_rate} for an underlying $16$-ASK. 
%As can be seen, for a wide range of SNRs, the GMI ($\rateHDD$) of the shaped constellation is very close to the MI ($\mI_{\mathsf{HSD}}$) of the underlying channel. 
As can be seen, the achievable information rate for the shaped constellation is significantly better than that of the uniform constellation. For an spectral efficiency of $3$ bits per channel use, the shaping gain is around $2$ dB. Table~\ref{T1} summarizes the shaping gain for different modulation orders and spectral efficiencies. 
In Fig.~\ref{FF2}, we depict the achievable rate $\rateHDD$ of the shaped constellation for different modulation orders. The figure shows that using the shaping described in this section, the CM system with HDD can operate at a roughly constant gap (around $1.9$ dB) to the capacity of the soft-decision AWGN channel.

 %This motivates us to design a binary code with hard-decoding for the hard-detection channel which will be explained in Section~\ref{coded_modulation}.

\section{Coded Modulation Scheme with PAS and Staircase Codes}\label{coded_modulation}

In this section, we apply PAS \cite{georg_tcom} to HDD and staircase codes. The CM scheme using PAS is depicted in Fig.~\ref{figsystemmodel}. 
As observed in \cite{georg_tcom},  $P_X^{{\lambda ^*}}$ is symmetric, i.e., $P_X^{{\lambda ^*}}\left( x \right) = P_X^{{\lambda ^*}}\left( { - x} \right)$ for $x \in {\cal X}$. Therefore, the random variable $X$ can be factorized as 
\begin{align}
\label{Amplitude_sign}
X = A \cdot S
\end{align}
where $A\triangleq |X|$ and $S \triangleq \mathsf{sign}(X)$ are the absolute value and sign of the RV $X$, respectively. It results that $S$ is uniformly distributed, 
\begin{align}
\label{Amplitude1}
P_S\left( 1 \right) = P_S\left( { - 1} \right)=\frac{1}{2}
\end{align}	
while the distribution of  $A$ satisfies
\begin{align}
\label{Amplitude}
P^{\lambda^{*}}_{A}\left( a \right) = 2P_X^{{\lambda ^*}}\left( {a} \right)
\end{align}
where $a \in \mathcal{A}\triangleq \{1,...,2^m-1\}$.

The idea of the PAS \cite{georg_tcom} is then to split the information sequence $\u$ into two sequences $\us$ and $\ua$. The sequence $\ua$ is used to generate a sequence of amplitudes $a_1,\ldots,a_n$ with the desired distribution. The binary image of the amplitudes and the remaining information bits, i.e., those belonging to $\us$, are then encoded using a binary code with a systematic encoder. The parity bits generated by the systematic encoder and $\us$ are used to generate $n$ sign labels $s_1,\ldots, s_n$. Assuming uniform distribution of the information bits and since the parity bits at the output of the encoder tend to be uniformly distributed as well, the sign labels closely mimic the desired (uniform) distribution \eqref{Amplitude1}.
A detailed description  of the CM scheme (see Fig.~\ref{figsystemmodel}) is provided in the following. 
	
\subsection{Amplitude Shaping}

Let $\u=(u_1,...,u_{k})$ be the information sequence of length $k$ bits, $u_i \in \{0,1\}$. Information bits are modeled as uniformly-distributed i.i.d. RVs. %One can see the uniform source in Fig.~\ref{figsystemmodel}, as the output of the source encoder. 
The vector $\u$ is split into two vectors $\us$ and $\ua$, of lengths $\gamma n$ and $k-\gamma n$, respectively, where $\gamma$ is a tuning parameter whose meaning will become clear later and it is assumed that $\gamma n \in \Np$. Vector $\ua$ is used to generate a sequence of amplitudes $\a=(a_1,...,a_n)$ with distribution $P^{\lambda^{*}}_{A}$ through a shaping block. The binary interface which generates the sequence of amplitudes with a given distribution (in our case $P^{\lambda^{*}}_{A}$) from a uniformly distributed input is called the \emph{distribution matcher}. The distribution matchers proposed in the literature can be categorized in two groups, variable-length \cite{kschischang1993optimal,Boc11,shape2,shape3} and fixed-length \cite{shapingGeorg} distribution matchers. To limit error propagation, we consider fixed-length distribution matching. In particular, we use the  \emph{constant composition distribution matching} (CCDM) method proposed in \cite{shapingGeorg}. This distribution matching algorithm uses arithmetic coding to generate the output amplitudes in an online fashion. Hence, there is no need for a lookup table as required by other algorithms. We refer the interested reader to \cite[Sec.~V]{georg_tcom} and \cite{shapingGeorg} for more details. The rate of the shaping block is $(k-\gamma n)/n$ and it approaches $\ent(A)$ for large $n$. We remark that the degree of parallelism in implementing the CCDM can be increased using the product distribution matcher \cite{bocherer2017highspeed} or the streaming distribution matcher \cite{bochererECOChighspeed}.

%\begin{table}[t]
%	%\renewcommand{\tabcolsep}{0.15cm}
%	%\vspace{-0.5cm}
%	\caption {Binary Gray mapping of the amplitudes for the 8-ASK modulation} \label{T2}
%	\vspace{-0.25cm}
%	\begin{center}
%	%	\scriptsize
%		\vspace{-2ex}
%		\begin{center}\begin{tabular}{ccccc}
%				\arrayrulecolor{black}\hline
%				\toprule
%				
%				
%				amplitude & 1  & 3 & 5 & 7 \\
%				\midrule
%				label & 00  & 01 & 11 & 10 \\
%				\hline
%				\toprule                			
%			\end{tabular} \end{center}
%		\end{center}
%		\vspace{-0.5cm}
%	\end{table}

\subsection{Amplitude-to-Bit Mapping}\label{mapping}

To reduce the number of bit errors associated with a symbol error, we consider the BRGC labeling. %Table~\ref{T2} summarizes the BRGC labeling used for the amplitudes of the $8$-ASK modulation. 
In the mapping part, for an $M$-ASK modulation with $m=\text{log}_2 M$ bits per symbol, the sequence of amplitudes $a_1,...,a_n$ is transformed into a sequence of bits using the mapper $\Phi_{\mathrm{ab}}$. We label each of the amplitudes $a_i$ with $m-1$ bits using the BRGC labeling to construct a binary string $\b_i\triangleq\b(a_i)= (b^{i}_1,\cdots,b^{i}_{m-1})$. The sequence $\b=(\b_1,\ldots,\b_n)$ is of length $\ell=n(m-1)$ bits.

\subsection{Encoding}\label{encoding}

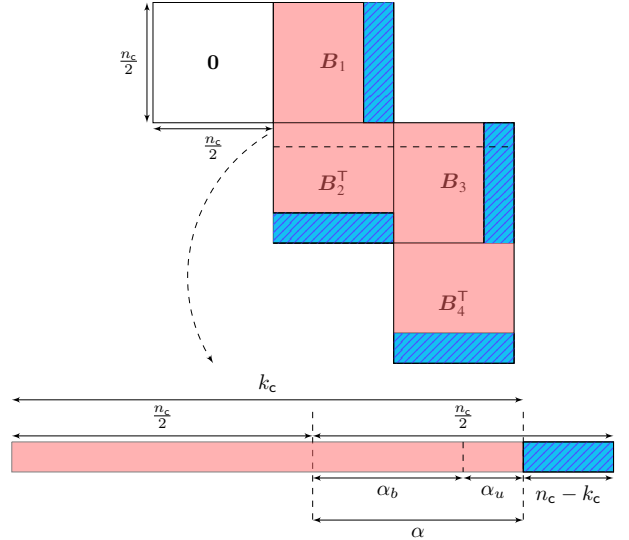
\begin{figure}[t]
\begin{center}
	\scalebox{0.8}{
		\begin{tikzpicture}[>=latex']	
		%%%%%%%  Staircase structure		
		\draw[black] (1,1) -- (1,-1) -- (-1,-1) -- (-1,1) -- (1,1); %square around the origin
		\draw[black] (1,1) -- (3,1) -- (3,-1) -- (1,-1) -- (1,1); %square around the origin		
		\draw[black] (1,-1) -- (1,-3) -- (3,-3) -- (3,-1); %square around the origin	
		\draw[black] (3,-1) -- (5,-1) -- (5,-3) -- (1,-3); %square around the origin
		\draw[black] (3,-3) -- (3,-5) -- (5,-5)  -- (5,-3); %square around the origin				
		
		%\draw [<->] (-1,1.1) to node[above] {$2m-r$} (2.45,1.1);
		%\draw [<->] (2.55,1.1) to node[above] {$r$} (3,1.1);
		\draw [<->] (-1,-1.1) to node[below] {$\frac{\nc}{2}$} (1,-1.1);
		\draw [<->] (-1.1,-1) to node[left] {$\frac{\nc}{2}$} (-1.1,1);
		
		%%%%%%%

		\draw [solid] (2.5,1) -- (2.5,-1);
		\draw [solid] (1,-2.5) -- (3,-2.5);        
		\draw [solid] (4.5,-1) -- (4.5,-3);
		\draw [solid] (3,-4.5) -- (5,-4.5);  
		
		%%%%%%%
		
	    \node (Zero) at (0,0) {$\bf 0$};
		\node (Zero) at (2,0) {$\Bmat_\text{1}$};		
		\node (Zero) at (2,-2) {$\Bmat_\text{2}^\rmT$};
		\node (Zero) at (4,-2) {$\Bmat_\text{3}$};		
		\node (Zero) at (4,-4) {$\Bmat_\text{4}^\rmT$};

		%%%%%%%
		\draw[black,fill=red!60,opacity=0.5] (1,1) -- (2.5,1) -- (2.5,-1) -- (1,-1) -- (1,1); %square around the origin
		\draw[black,fill=cyan!70,opacity=1] (2.5,1) -- (2.5,-1) -- (3,-1) -- (3,1); %square around the origin 
		\draw[pattern=north east lines, pattern color=blue!70] (2.5,1) -- (2.5,-1) -- (3,-1) -- (3,1); %square around the origin		               
		\draw[black,fill=red!60,opacity=0.5] (1,-1) -- (1,-2.5) -- (3,-2.5) -- (3,-1); %square around the origin
        \draw[black,fill=cyan!70,opacity=1] (1,-2.5) -- (3,-2.5) -- (3,-3) -- (1,-3); %square around the origin 
         \draw[pattern=north east lines, pattern color=blue!70] (1,-2.5) -- (3,-2.5) -- (3,-3) -- (1,-3); %square around the origin         
        \draw[black,fill=red!60,opacity=0.5] (3,-3) -- (4.5,-3) -- (4.5,-1) -- (3,-1); %square around the origin
         \draw[black,fill=cyan!70,opacity=1] (4.5,-3) -- (4.5,-1) -- (5,-1) -- (5,-3); %square around the origin 
         \draw[pattern=north east lines, pattern color=blue!70] (4.5,-3) -- (4.5,-1) -- (5,-1) -- (5,-3); %square around the origin          
        \draw[black,fill=red!60,opacity=0.5] (5,-3) -- (5,-4.5) -- (3,-4.5) -- (3,-3); %square around the origin
         \draw[black,fill=cyan!70,opacity=1] (5,-4.5) -- (5,-5) -- (3,-5) -- (3,-4.5); %square around the origin  		
         \draw[pattern=north east lines, pattern color=blue!70] (5,-4.5) -- (5,-5) -- (3,-5) -- (3,-4.5); %square around the origin 		
		%%%%%%%
		\draw [dashed] (1,-1.4) -- (5,-1.4);
		
        \draw [dashed,->] (0.9,-1.2) to [bend right=45] (0,-5); 
		
		\end{tikzpicture}}
	
\vspace{0.05cm}
	
	\scalebox{0.8}{
		\begin{tikzpicture}[>=latex']	
		
		%%%%%%% Component code	
		%\node[] at (1,-6.9) {Component code: $n=2m$, $k=2m-r$};
		\draw [<->] (-5,-5.9) to node[above] {$\frac{\nc}{2}$} (0,-5.9);	
		\draw [<->] (0,-5.9) to node[above] {$\frac{\nc}{2}$} (5,-5.9);	
		\draw [dashed] (0,-5.5) -- (0,-7.3);					
		\draw[black,fill=red!60,opacity=0.5] (-5,-6) -- (3.5,-6) -- (3.5,-6.5) -- (-5,-6.5) -- (-5,-6); 		
		%\node[] at (1,-5.8) {\text{Information}};		
		\draw[black,fill=cyan!70,opacity=1] (3.5,-6) -- (5,-6) -- (5,-6.5) -- (3.5,-6.5) -- (3.5,-6);
		\draw[pattern=north east lines, pattern color=blue!70] (3.5,-6) -- (5,-6) -- (5,-6.5) -- (3.5,-6.5) -- (3.5,-6);		
		%\node[] at (4.1,-6.25) {\text{Redundancy}};		    		
		%%%%%%%
		\draw [<->] (0,-6.6) to node[below] {$\alpha_b$} (2.5,-6.6);		
		\draw [<->] (2.5,-6.6) to node[below] {$\alpha_u$} (3.5,-6.6);	
		\draw [<->] (3.5,-6.6) to node[below] {$\nc-\kc$} (5,-6.6);	
		
		\draw [dashed] (3.5,-5.5) -- (3.5,-7.3);	
		\draw [<->] (0,-7.3) to node[below] {$\alpha$} (3.5,-7.3);	
				
		\draw [<->] (-5,-5.3) to node[above] {$\kc$} (3.5,-5.3);		
		\draw [dashed] (2.5,-6) -- (2.5,-6.5);		
		\end{tikzpicture}}
\end{center}
	\vspace{-3ex} 
	\caption{Code array of a staircase code. The red colored area shows the information bits while the area shown with blue hatches corresponds to the parity bits.}
	\label{staircase_code}
\vspace{-2ex}
\end{figure}
	
The sequences $\b$ and $\us$ are multiplexed and encoded by a binary linear block code with a systematic encoder. %explained in the following. %We denote by $\p=(p_1,...,p_{n(1-\gamma)})$ the vector of parity bits, of length $n(1-\gamma)$, at the output of the binary code. As shown in \cite[Sec.~IV.A]{georg_tcom}, for systematic codes the distribution of the information bits are preserved while the parity bits have the uniform distribution. 
As binary linear block codes, we use staircase codes with Bose-Chaudhuri-Hocquenghem (BCH) codes as component codes. In particular, we consider $(\nc,\kc)$ systematic BCH component codes of code length $\nc$ and dimension $\kc$. Staircase codes can be decoded iteratively using bounded-distance decoding (BDD) for the decoding of the component codes and can provide a $0.42$ dB coding gain over the best known code from the ITU-T G.975.1 recommendation \cite{staircase_frank}. Let $\mathcal{C}$ be an $(\nc,\kc)$ shortened BCH code constructed over the Galois field $\text{GF}(2^v)$ with (even) block length $\nc$ and information block length $\kc$ given by
	\begin{align}
	& \nc=2^v-1-s, \label{n_formula}\\
	& \kc=2^v-vt-1-s \label{k_formula}
	\end{align} 
where $s$ and $t$ are the shortening length and the error correcting capability of the code, respectively. A shortened BCH code is thus completely specified by the parameters $(v,t,s)$. A staircase code with $(\nc,\kc)$ component codes is defined as the set of all $\frac{\nc}{2} \times \frac{\nc}{2}$ matrices $\Bmat_i$, $i=1,2,\ldots$, such that each row of the matrix $[\Bmat_{i-1}^\rmT,\Bmat_i]$ is a valid codeword of $\mathcal{C}$. Fig.~\ref{staircase_code} shows the code array of a staircase code. The red parts correspond to the information bits while the parity bits are shown with blue hatches.

\begin{table*}[ht]
	\renewcommand{\tabcolsep}{0.15cm}
	\caption {Parameters of the designed staircase codes for $v=10$, $t=3$, and $16$-ASK modulation} \label{table:feasibleparam}
	\vspace{-0.25cm}
	\begin{center}
		%\scriptsize
		\vspace{-2ex}
		\begin{center}\begin{tabular}{ccccccccccccc}
				\arrayrulecolor{black}\hline
				\toprule

				$\gamma$ & 0.75 & 0.6907 & 0.6809 & 0.5946 & 0.5238 & 0.4444 & 0.4 &  0.3617 & 0.3023 & 0.2308 & 0.1429 & 0\\
				\midrule
				$s$ & 63 & 247 & 271 & 431 & 519 & 591 & 623 & 647 & 679 & 711 & 743  & 783 \\
				\midrule				
				$\nc$ & 960 & 776 & 752  &  592 & 504 & 432 & 400 & 376 & 344 & 312 & 280 & 240 \\				
				\midrule				
				$\kc$ & 930 & 746 & 722 & 562 & 474 & 402 & 370 & 346 & 314 & 282 & 250 & 210 \\								
				\midrule				
				$n$ & 57600 & 37636 & 35344 & 21904 & 15876 & 11664 & 10000 & 8836 & 7396 & 6084 & 4900 & 3600\\
				\midrule				
				$\alpha_u$ & 90 & 67 & 64 & 44 & 33 & 24 &  20 & 17 & 13 & 9 & 5 & 0\\			
				\midrule				
				$\Rs$ & 0.9375 & 0.9227 & 0.9202 & 0.8986 & 0.8810 & 0.8611 & 0.85 & 0.8404 & 0.8256 & 0.8077 & 0.7857 & 0.75\\		
				
				\hline
				\toprule                			
			\end{tabular} \end{center}
		\end{center}
		\vspace{-0.5cm}
	\end{table*}

\begin{table}[!t]
	\renewcommand{\tabcolsep}{0.15cm}
	\caption {Parameters of the designed staircase codes for $v=10$, $t=3$, and $8$-ASK modulation} \label{table:feasibleparam_8askmod}
	\vspace{-0.25cm}
	\begin{center}
		%\scriptsize
		\vspace{-2ex}
		\begin{center}\begin{tabular}{ccccccccc}
				\arrayrulecolor{black}\hline
				\toprule

				$\gamma$ & 0.8125 & 0.75 & 0.6 & 0.5 & 0.4444 & 0.25 &  \\
				\midrule
				$s$ & 63 & 303 & 573 & 663 & 669 & 783 & \\
				\midrule				
				$\nc$ & 960 & 720 &  450 & 360 & 324 & 240 & \\				
				\midrule				
				$\kc$ & 930 & 690 & 420 & 330 & 294 & 210 &  \\								
				\midrule				
				$n$ & 76800 & 43200 & 16875 & 10800 & 8748 & 4800 &  \\
				\midrule				
				$\alpha_u$ & 130 & 90 & 45 & 30 & 24 &  10 &  \\			
				\midrule				
				$\Rs$ & 0.9375 & 0.9167 &  0.8667 & 0.8333 & 0.8148 & 0.75 & \\		
			
				\hline
				\toprule                			
			\end{tabular} \end{center}
		\end{center}
		\vspace{-0.5cm}
	\end{table}
	
%To incorporate the staircase code in the proposed coded modulation, we should design $n$ and $\gamma$ based on the parameters of the component code. 

The rate of the staircase code with $(\nc,\kc)$ BCH component codes is \begin{align}
\label{st_rate}
{\Rs} = 1 - \frac{2(\nc-\kc)}{\nc}.
\end{align} 
For a $2^m$-ASK constellation, we assume a staircase code with code rate
\begin{align}
\label{componentcoderate}
\Rs = 1 - \frac{2(\nc-\kc)}{\nc} =  \frac{{m - 1 + \gamma }}{m}
\end{align} 
where $0 \le \gamma  < 1$ is a tuning parameter which can be used to select the rate of the staircase code and subsequently the spectral efficiency of the CM scheme.

For $(\nc,\kc)$ component codes, each matrix $\Bmat_i$ of the staircase code array contains $\alpha= (\kc-\nc/2)$ information bits per row, i.e., each matrix $\Bmat_i$ contains a total of $\alpha(\nc/2)$ information bits (see Fig.~\ref{staircase_code}). Consequently, we parse the sequences $\b$ and $\us$ into blocks of length $n$ and $\gamma n$, respectively, such that
\begin{align}
n=\frac{\alpha\cdot(\nc/2)}{m-1+\gamma}
\end{align}
where the parameters $(\nc,\kc)$, and hence $\gamma$, are chosen such that $n$ and $\gamma n$ are integers. Using \eqref{componentcoderate}, it follows that the number of parity bits in each row of $\Bmat_i$, $\nc-\kc$, is
\begin{align}
\label{eq:ncmkc}
\nc-\kc&=\alpha\left(\frac{1}{R_s}-1\right)&\nonumber\\
&=\frac{n(m-1+\gamma)}{\nc/2}\left(\frac{m}{m-1+\gamma}-1\right)\nonumber\\
&=\frac{n(1-\gamma)}{\nc/2}.
\end{align}

The bits $\b_{(i-1)\cdot n+1}, \b_{(i-1)\cdot n+2},\ldots, \b_{(i-1)\cdot n+n}$ and ${\us}_{(i-1)\cdot \gamma n+1},\ldots {\us}_{(i-1)\cdot \gamma n+n}$, $i=1,2,\ldots$, are then placed in $\Bmat_{i}$. In particular, let $\tilde{\b}_i$ be the string of bits obtained concatenating $\b_{i\cdot n+1}, \b_{i\cdot n+2},\ldots, \b_{i\cdot n+n}$ and $\tus_i$ the string of bits obtained concatenating ${\us}_{i\cdot \gamma n+1},\ldots {\us}_{i\cdot \gamma n+n}$. Then, $\b_i$ and $\tus_i$ are divided into $\nc/2$ equal parts, $\b_{i,1},\ldots,\b_{i,\nc/2}$ and $\tus_{i,1},\ldots,\tus_{i,\nc/2}$, of lengths $\alpha_b=n(m-1)/(\nc/2)$ and $\alpha_u=\gamma n / (\nc/2)$, respectively, and the vector $(\b_{i,j},\us_{i,j})$ is placed in row $j$ of matrix $\Bmat_{i+1}$. Note that $(\nc,\kc)$ and $\gamma$ must be chosen such that
\begin{align}
\frac{n}{\nc/2}&\in \N \label{eq:cond2}\\
\frac{\gamma n}{\nc/2}&\in\Np.\label{eq:cond3}
\end{align}

Consider for simplicity the code array blocks $\Bmat_1$ and $\Bmat_2$. In this case, vector $(\b_{0,j},\us_{0,j})$ is placed in row $j$ of $\Bmat_1$, for $j=1,\ldots,\nc/2$, and vector $(\b_{1,j},\us_{1,j})$ is placed in row $j$ of $\Bmat_2$, for $j=1,\ldots,\nc/2$. Similarly, the rows of matrices $\Bmat_i$ for $i>2$ are filled. Then, encoding of the staircase code is performed as usual, by row/column encoding, using the $(\nc,\kc)$ BCH component code (see Fig.~\ref{staircase_code}). We denote by $\p_i$, $i=1,\ldots,\nc(\nc-\kc)/2$, the parity bits of matrix $\Bmat_i$ (shown in blue hatches in Fig.~\ref{staircase_code}) resulting from the encoding process, and by $\p_{i,j}$, $j=1,\ldots,\nc/2$ the parity bits in the $j$-th row of $\Bmat_i$. Note that each vector $\p_{i,j}$ is of length $\nc-\kc$ bits.

If the ASK constellation is fixed, to achieve different spectral efficiencies, the code rate of the staircase code must be changed. For a given modulation order (given $m$), one can find different feasible solutions for $(v, t, s)$ that lead to a value of $\gamma$ that satisfies \eqref{eq:cond2}, \eqref{eq:cond3}, and $n\in\N$. 

In Table~\ref{table:feasibleparam} and Table~\ref{table:feasibleparam_8askmod}, some feasible values for $s$ (and therefore $\gamma$), the parameters $(\nc,\kc)$ for the resulting BCH component codes, $n$, and the rate of the staircase code, $\Rs$, are summarized for $16$-ASK modulation ($m=4$) and $8$-ASK modulation ($m=3$), for $v=10$ and $t=3$.

\subsection{Bit-to-Sign Mapping}\label{demapping}

In the demapping block, for each code array block $\Bmat_i$, the parity bits of $\Bmat_i$ are multiplexed with $\tus_i$ to generate, through the bit-to-sign mapper $\Phi_{\mathrm{bs}}$, the sequence of signs $\s=(s_1,\ldots,s_n)$, which will be used as the signs of the amplitudes $a_1,\ldots,a_n$. In particular, we combine the block $\p_{i,j}$ of $\nc-\kc$ parity bits (corresponding to the parity bits of row $j$ of $\Bmat_i$) with the $\gamma n /(\nc/2)$ information bits $\tus_ {i,j}$ to form the vector $\t_{i,j}=(\p_{i,j},\tus_ {i,j})$, of length $\nc-\kc+\gamma n /(\nc/2)=n/(\nc/2)$, where we used \eqref{eq:ncmkc}. Then, the binary vector $\t_i=(\t_{i,1},\ldots,\t_{i,\nc/2})=(t_1,\ldots,t_n)$, of length $n$ bits, is used to form the sequence of $n$ signs $\s$ by setting
$s_i = 2{t_i} - 1$.
Recalling that the distribution of the signs closely mimics the uniform one,
according to \eqref{Amplitude_sign}--\eqref{Amplitude}, the element-wise multiplication of $(a_1,\ldots,a_n)$ by $(s_1,\ldots,s_n)$ generates a sequence $\x=(x_1,\ldots,x_n)$ with the desired distribution.

Finally, the sequence $\x$ is scaled by $\Delta$ and transmitted over the fiber-optic channel (FOC).

%We assume that the bit corresponding to the sign of each symbol is attached to the label of the underlying amplitude as the most significant bit, giving the label of each symbol. %Table~\ref{tab:labelsASK} summarizes the labels for $8$-ASK. 
%As can be seen, with this labeling the symbols and amplitudes follow the BRGC labeling. One can use these labels to compute the \GL{rate} from \eqref{GMI_rate}.
%\begin{table}
%	\renewcommand{\tabcolsep}{0.15cm}
%	\caption {BRGC mapping of the symbols for $8$-ASK modulation} \label{tab:labelsASK}
%	\vspace{-0.25cm}
%	\begin{center}
%		\scriptsize
%		\vspace{-2ex}
%		\begin{center}\begin{tabular}{ccccccccc}
%				\arrayrulecolor{black}\hline
%				\toprule
%				
%				
%				symbol & -7  & -5 & -3 & -1 & 1  & 3 & 5 & 7\\
%				\midrule
%				label & 010  & 011 & 001 & 000 & 100  & 101 & 111 & 110\\
%				\hline
%				\toprule                			
%			\end{tabular} \end{center}
%		\end{center}
%		\vspace{-0.5cm}
%	\end{table}

%%%%%%%%%%%%%%%%%%%%%%%%%%%%%%%%%%%%%%%%%%%%%%%%%%%%%%%%%%%%%%%%%%%%%%%%%%%%%%%%%%%%%%%%%

%%%%%%%%%%%%%%%%%%%%%%%%%%%%%%%%%%%%%%%%%%%%%%%%%%%%%%%%%%%%%%%%%%%%%%%%%%%%%%%%%%%%%%%%%%%%%%%

\section{Designing the Operating Point}\label{operating_point}

\subsection{The Effect of Shaping on the Operating Point}\label{shaping_operating}

For each staircase block $\Bmat_i$, the transmitted data is ${\us}_{(i-1)\cdot \gamma n+1},\ldots, {\us}_{(i-1)\cdot \gamma n+n}$ and the information bits in $\ua$, which are embedded in the amplitudes $a_{(i-1)\cdot n+1}, a_{(i-1)\cdot n+2},\ldots, a_{(i-1)\cdot n+n}$.
%$A_{(i-1)\cdot n+1}, A_{(i-1)\cdot n+2},\ldots, A_{(i-1)\cdot n+n}$. 
Therefore, the rate is $\rmR=\ent(A) + \gamma$ [bits/channel use].
An arbitrarily low error probability can be achieved if
\begin{align}
\label{operating_cond}
\rmR=\ent(A) + \gamma<\rateHDD.
\end{align}

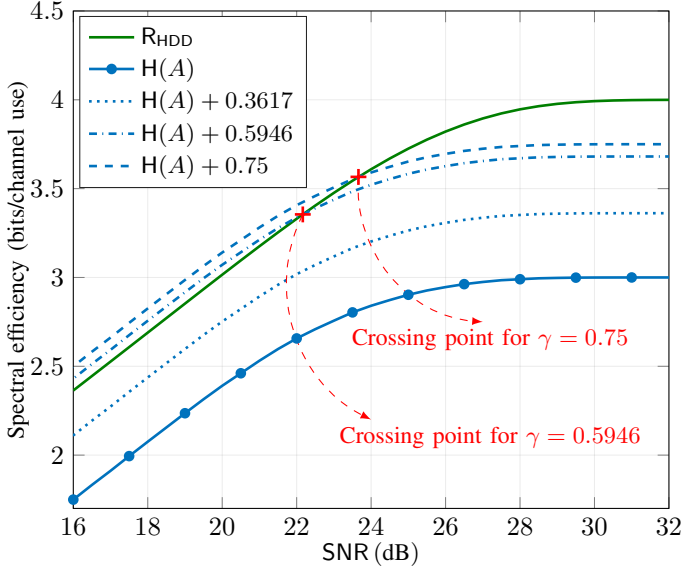
\begin{figure}[!t] \centering 
	\setlength\figureheight{8.2cm}
	\setlength\figurewidth{9.5cm}  \input{operating_point_update_new_map_newgeorg.tikz} 
	\vspace{-5ex}
	\caption{An example for determining the feasible SNR region depending on the selected $\gamma$ corresponding to some of the designed codes for $16$-ASK.} \vspace{-2ex}
	\label{feasible_region} 
\end{figure}

The crossing point between the curves $\ent(A) + \gamma$ and $\rateHDD$ determines the optimal operating point \cite{georg_tcom}, which could be achieved by a capacity achieving code with infinitely large block length. In Fig.~\ref{feasible_region}, we depict $\rateHDD$ and $\rmR=\ent(A) + \gamma$ for different values of $\gamma$ corresponding to some of the designed codes summarized in Table~\ref{table:feasibleparam}. As can be seen, for $\gamma=0$ and $\gamma=0.3617$, all points on the curve $\ent(A)+\gamma$ corresponding to the SNRs in the displayed SNR range are achievable (the curve $\ent(A) + \gamma$ is below $\rateHDD$). However, for $\gamma=0.5946$ and $\gamma=0.75$ only the points on the curve  $\ent(A)+\gamma$ corresponding to SNRs larger than $22.16$ dB and $23.66$ dB, respectively, are achievable. Note that below these values, \eqref{operating_cond} is not satisfied. We remark that, in principle, the points on the curve $\ent(A)+\gamma$ in the feasible SNR region are only achievable using codes with infinite code length. In practice, codes operate at finite length. For finite code length and a given $\gamma$, one can simulate the performance of the designed system and find the minimum SNR in the feasible SNR region required to achieve the desired block error probability.

\subsection{Parameters of the Staircase Code}\label{code_operating}

For a given ASK constellation and a given spectral efficiency, i.e., fixed $\gamma$ and hence fixed staircase code rate $\Rs$, one can find several component BCH codes $(v,t,s)$ that satisfy \eqref{st_rate} and conditions \eqref{eq:cond2}, \eqref{eq:cond3}, and $n\in\N$, i.e., one can find several staircase codes that yield the desired spectral efficiency. Among them, we may then choose the one that yields the best decoding threshold, i.e., the lowest SNR at which the probability of error goes to zero for infinite block length, using the density evolution derived in \cite{christian1}. This approach leads, for each spectral efficiency, to the best possible staircase code and therefore to the best performance. However, changing code for each spectral efficiency may not be desirable in practice. An alternative approach is to fix the parameters $(v,t)$ of the component BCH codes and then tune the shortening parameter $s$, which leads to different $\gamma$ and thus to different spectral efficiencies. With this approach one can cover a wide range of spectral efficiencies with a single staircase code.
 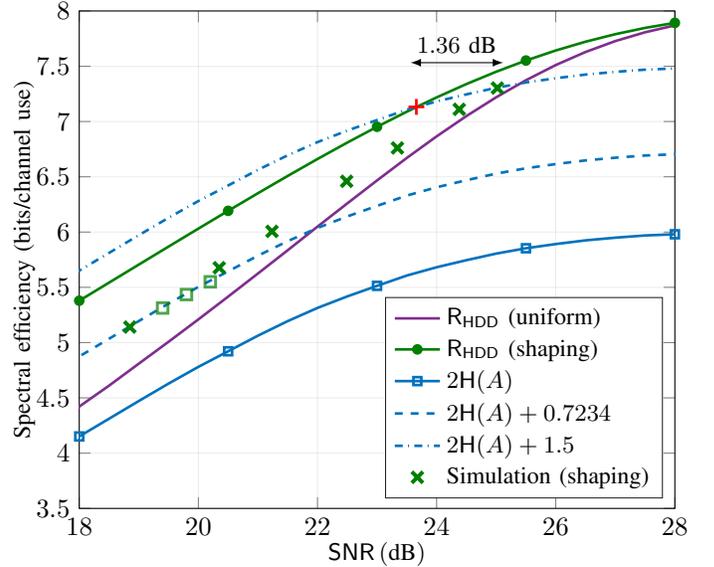
\begin{figure}[!t] \centering 
	\setlength\figureheight{8.2cm}
	\setlength\figurewidth{9.5cm}  \input{F1_16ASK_illustration_new_MAP_QAMupdate_georgrate.tikz} 
	\vspace{-5ex}
	\caption{Simulation results of the CM scheme for $256$-QAM using the staircase codes summarized in Table~\ref{table:feasibleparam}, and comparison with the corresponding achievable information rates.} \vspace{-1ex}
	\label{16_ASK_sim} 
\end{figure}

Note that since one can see square QAM constellations as the Cartesian product of two ASK constellations, the design of the PAS scheme described above readily extends to QAM constellations. In this case, the rate is twice that of the scheme with ASK constellation, i.e., $\rmR=2\ent(A) + 2\gamma$.
In Fig.~\ref{figsystemmodel}, the dashed block multiplexes two ASK modulated sequences corresponding to generate the real and imaginary components of the QAM constellation.

\section{Numerical Results}\label{sec:simulation}
We assess the performance of the CM scheme with PAS and HDD using the codes summarized in Table~\ref{table:feasibleparam} and Table~\ref{table:feasibleparam_8askmod} for transmission using square QAM constellations. The target block error probability is set to $\Pe=10^{-3}$. Here, by block error probability we mean the probability that a block of bits of $\ua$ and $\us$ corresponding to a staircase block is in error. By means of simulations, we find the minimum SNR for which the target $\Pe$ is achieved. At the receiver, a symbol-wise MAP detector is used and the decoding of the staircase code is performed using a sliding-window decoder with window size of $7$ staircase blocks and a maximum of $8$ decoding iterations within the window based on the BDD with extrinsic message passing \cite[Algorithm 1]{h_braided}.

To minimize the number of the operating modes, we consider the use of a single staircase code (according to the second approach described in Section~\ref{code_operating}). In particular, we consider a staircase code with BCH component codes with parameters $(v,t)=(10,3)$. To achieve different code rates, i.e., different spectral efficiencies, we then find suitable shortening parameters $s$ which lead to a value of $\gamma$ that satisfies (\ref{eq:cond2}), (\ref{eq:cond3}). Some of the values of $s$ and $\gamma$ are summarized in Table~\ref{table:feasibleparam} and Table~\ref{table:feasibleparam_8askmod}.

In Fig.~\ref{16_ASK_sim}, we plot the \emph{practical operating points} of the PAS scheme for $256$-QAM (green crosses), corresponding to the minimum SNR required to achieve $\Pe=10^{-3}$ using the optimal distribution $P_X^{{\lambda ^*}}$ for an underlying $16$-ASK constellation (the $256$-QAM is then obtained as the Cartesian product of two ASKs with distribution $P_X^{{\lambda ^*}}$). The shortening parameters are $63$, $274$, $431$, $519$, $591$, $623$, and $647$ (starting from the cross at the top right). The corresponding values of $\gamma$ are given in the first row of Table~\ref{table:feasibleparam}. Note that we can vary the spectral efficiency from $5.14$ to $7.3$ bits/channel use using a single code and decoder by simply changing the shaping distribution and shortening of the component code. Remarkably, the rates achieved by the probabilistically-shaped CM scheme are larger than the achievable rate of the CM scheme with uniform distribution (purple curve).

As an example, by finding the crossing point between the achievable rate curve and $\rateHDD$ for $\gamma=0.75$ ($s=63$), marked with a red plus sign in the figure, one can see that for all SNRs larger than $23.66$ dB reliable transmission can be achieved. However, when using a practical, finite-length staircase code, one should back-off roughly $1.36$ dB to meet the target $\Pe$.  We also remark that all points on the curves $2\ent(A) + 2\gamma$ with SNR larger than the minimum required SNR can also meet the target performance. However, since by increasing the SNR the curve $2\ent(A) + 2\gamma$ flattens out, at some point one needs to switch to another code rate (another $\gamma$) in order to operate as close as possible to the curve $\rateHDD$. For instance, for $\gamma=0.3617$ ($s=647$), the other possible operating points are shown with green squares. For $20.35$ dB, one should switch to $\gamma=0.4$ ($s=623$) to approach $\rateHDD$ more closely. Furthermore, to have a fine granularity for the achievable spectral efficiencies, we can consider several operating points on the same achievable rate curve, e.g., squares on the achievable rate curve $2\ent(A)+0.7234$ corresponding to $\gamma=0.3617$. 

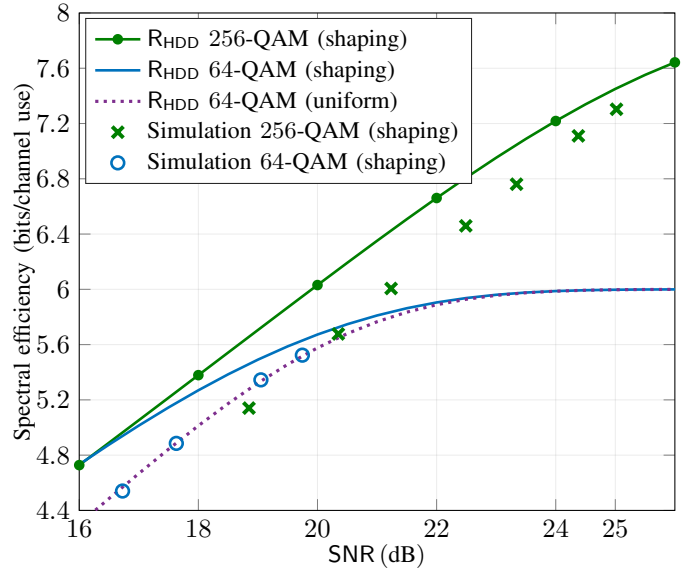
\begin{figure}[!t] \centering 
	\setlength\figureheight{8.2cm}
	\setlength\figurewidth{9.5cm}  \input{F2_16ASK_illustration_new_MAP_QAMupdate_georgrate.tikz} 
	\vspace{-5ex}
	\caption{Simulation results of the CM scheme for $256$-QAM and $64$-QAM using the staircase codes summarized in Table~\ref{table:feasibleparam} and the first four columns of Table~\ref{table:feasibleparam_8askmod}, and comparison with the corresponding achievable information rates.}  \vspace{-2ex}
	\label{8_ASK_sim} 
\end{figure}

\begin{figure}[!t] \centering 
    \includegraphics[width=\linewidth]{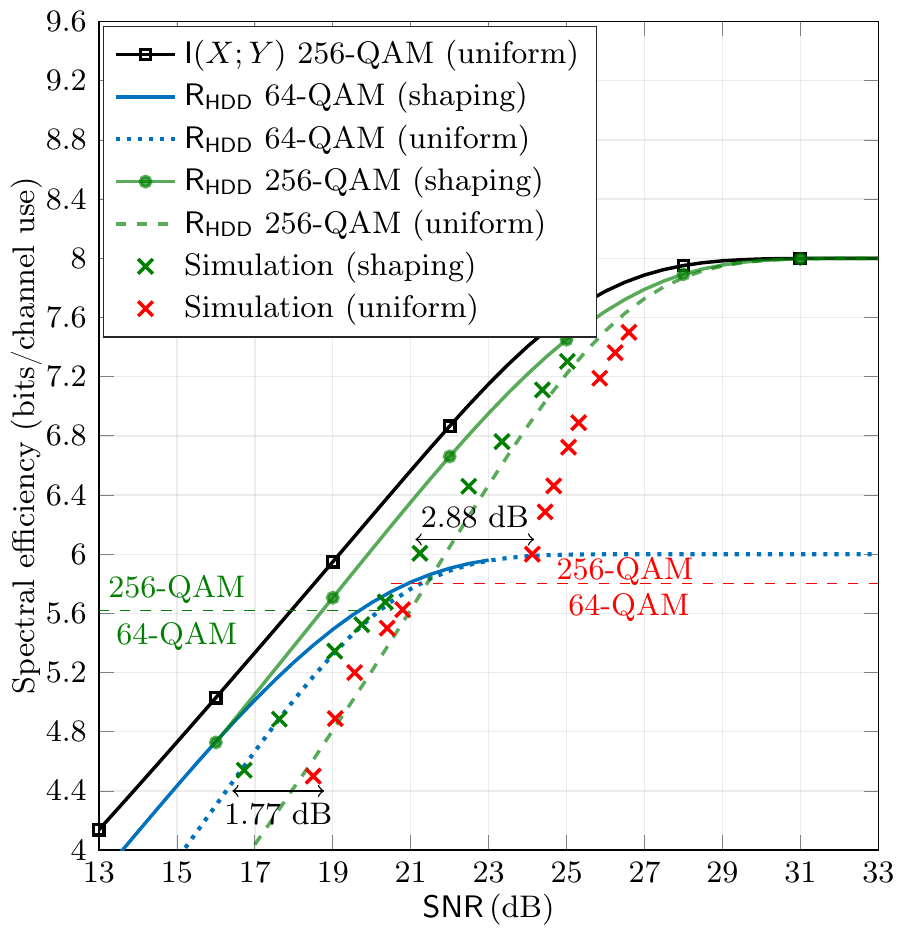}
	\vspace{-5ex}
	\caption{Performance of the probabilistically-shaped CM scheme using $P_X^{{\lambda ^*}}$ and the CM scheme with uniform distribution.}  \vspace{-2ex}
	\label{sim_total} 
\end{figure}

%For a given modulation order, a range of spectral efficiencies can be covered by shortening a single BC component code, hence one need to change the modulation order which in turn leads to possibly different component codes (cf. Table~\ref{table:feasibleparam_8askmod} and Table~\ref{table:feasibleparam}). 
In Fig.~\ref{8_ASK_sim}, we plot the performance of the CM scheme with $256$-QAM modulation (green crosses) and $64$-QAM modulation (blue circles). For $256$-QAM, the shortening parameters are the same as in Fig.~\ref{16_ASK_sim}. For $64$-QAM, the shortening parameters are $63$, $303$, $573$, and $663$. %\RF{As can be seen, as a coincidence, the performance of the probabilistically-shaped CM scheme with $64$-QAM is approximately on top of the achievable rate curve of the scheme with uniform distribution}. 
We observe that the performance of the CM scheme with $64$-QAM for spectral efficiencies below $5.52$ bits/channel use is better than that of the scheme with $256$-QAM, showing that, depending on the spectral efficiency region, one should change the underlying constellation to achieve better performance. Also, interestingly, the performance for $64$-QAM is closer to the corresponding achievable rate curve than for $256$-QAM. We believe that this is due to the fact that for a given spectral efficiency that is achievable with both $64$-QAM and $256$-QAM, the codes used for $64$-QAM have higher rate than those for $256$-QAM and it is well known that high-rate staircase codes perform better than lower-rate staircase codes\cite{staircase_frank,Smith_2012}. If the complexity of the system is of significant importance, one may disregard the additional gain of switching from $256$-QAM to $64$-QAM, and consider only $256$-QAM.  

Finally, in Fig.~\ref{sim_total} we compare the performance of the probabilistically-shaped CM scheme with $64$-QAM and $256$-QAM (green crosses) with that of the scheme with uniform distribution (red crosses). For the sake of comparison, we also depict the achievable rate $\rateHDD$ of the CM scheme with the shaped distribution and the mutual information of the scheme with uniform distribution and symbol-wise SDD, given by $\mathsf{I}(X;Y)$ in \eqref{eq:MI}, both for $256$-QAM.
For the system with uniform distribution, the shortening parameters are $63$, $271$, $431$, $591$, $647$, $711$, $743$, and $783$ for $256$-QAM, and $63$, $303$, $573$, $669$, and $783$ for $64$-QAM. The probabilistic amplitude shaping CM scheme achieves significantly better performance compared to that of the baseline scheme with uniform distribution. Gains up to $2.88$ dB and $1.77$ dB are achieved for $256$-QAM and $64$-QAM, respectively. Furthermore, the probabilistically-shaped CM scheme achieves performance within $0.57$--$1.44$ dB of the corresponding achievable rate for a wide range of spectral efficiencies.

%operates roughly in a gap ranging from $0.58$--$1.45$ dB to the ${I}_{\mathsf{HDD}}^{\mathsf{GMI}}$ while the same gap (with respect to ${I}_{\mathsf{HDD}}^{\mathsf{GMI}}$ with uniform distribution) for the uniform system varies between $0.55$--$2.28$ dB.  

\section{Conclusion}\label{conclusion}

We applied probabilistic amplitude shaping to binary staircase codes with hard decision decoding for high-speed fiber-optic communications. We optimize the input distribution to maximize a relevant achievable rate of the CM system with PAS and bit-wise HDD. Outstandingly, the performance of the CM scheme with PAS is significantly better than that of the standard CM scheme with uniform distribution. The probabilistically-shaped CM scheme achieves performance within $0.57$--$1.44$ dB of the achievable rate for a wide range of spectral efficiencies using only a single staircase code, which greatly reduces the decoder complexity.  

\section*{Acknowledgment}
The authors would like to thank Dr. Georg B\"ocherer for bringing to our attention the achievable rate for probabilistic amplitude shaping in \cite{bocherer2017} and useful comments.

\balance

\end{document}

%% file: MI_GMI_16ASK_without_tol_new_MAP_georgrate.tikz
\definecolor{mycolor1}{rgb}{0.00000,0.44700,0.74100}%
\definecolor{mycolor2}{rgb}{0.00000,0.44706,0.74118}%
\begin{tikzpicture}
\pgfplotsset{every tick label/.append style={font=\small}}
\begin{axis}[%
width=\figurewidth,
height=\figureheight,
at={(0\figurewidth,0\figureheight)},
scale only axis,
ylabel={\small Spectral efficiency (bits/channel use)},
xlabel={\small $\SNR\, (\textrm{dB})$},
ylabel style={yshift=-0.58cm},
xlabel style={yshift=0.2cm},
xmin=16,
xmax=30,
grid style={gray,opacity=0.15},
xmajorgrids,
ymin=1.5,
ymax=4.5,
ymajorgrids,
axis background/.style={fill=white},
legend style={at={(0.98,0.02)},anchor=south east,legend cell align=left,align=left,draw=white!15!black}
]

\addplot [color=black,solid,line width=1.0pt,,mark size=1.5pt,mark=square,mark repeat={4} ,mark options={solid}]
table[row sep=crcr]{%
	0	0.5\\
	0.5	0.542718601409599\\
	1	0.587818317346194\\
	1.5	0.635275697101155\\
	2	0.685052334875493\\
	2.5	0.737095848832961\\
	3	0.791341177455778\\
	3.5	0.847712127039064\\
	4	0.906123095650313\\
	4.5	0.96648089622865\\
	5	1.0286866043034\\
	5.5	1.0926373631491\\
	6	1.15822808981313\\
	6.5	1.22535303798124\\
	7	1.29390718678102\\
	7.5	1.36378743725932\\
	8	1.43489360958514\\
	8.5	1.50712924350316\\
	9	1.58040221195651\\
	9.5	1.65462516310201\\
	10	1.72971580931865\\
	10.5	1.8055970835281\\
	11	1.88219718352143\\
	11.5	1.95944952434599\\
	12	2.03729261745271\\
	12.5	2.11566989350116\\
	13	2.19452948368152\\
	13.5	2.27382397230579\\
	14	2.35351013136442\\
	14.5	2.43354864582107\\
	15	2.51390383667526\\
	15.5	2.59454338728602\\
	16	2.6754380771243\\
	16.5	2.75656152600264\\
	17	2.83788995090244\\
	17.5	2.91940193676406\\
	18	3.00107822200099\\
	18.5	3.08290149902681\\
	19	3.16485622972095\\
	19.5	3.24692847548876\\
	20	3.3291057413759\\
	20.5	3.41137683356277\\
	21	3.49373172947796\\
	21.5	3.5761614597214\\
	22	3.65865800096827\\
	22.5	3.74121417902728\\
	23	3.8238235812452\\
	23.5	3.90648047747976\\
	24	3.98917974890062\\
	24.5	4.07191682392081\\
	25	4.1546876206064\\
	25.5	4.23748849495864\\
	26	4.3203161945087\\
	26.5	4.40316781671055\\
	27	4.48604077166048\\
	27.5	4.56893274871298\\
	28	4.65184168660144\\
	28.5	4.73476574670804\\
	29	4.81770328916082\\
	29.5	4.90065285146676\\
	30	4.983613129418\\
	30.5	5.0665829600343\\
	31	5.14956130632863\\
	31.5	5.23254724370396\\
	32	5.31553994780929\\
	32.5	5.39853868370022\\
	33	5.48154279616551\\
	33.5	5.5645517010955\\
	34	5.64756487778109\\
	34.5	5.73058186204385\\
	35	5.81360224010802\\
	35.5	5.89662564313503\\
	36	5.97965174234899\\
	36.5	6.06268024468985\\
	37	6.1457108889372\\
	37.5	6.22874344225396\\
	38	6.31177769710469\\
	38.5	6.394813468508\\
	39	6.47785059158696\\
	39.5	6.56088891938526\\
	40	6.64392832092027\\
	40.5	6.72696867944758\\
	41	6.81000989091377\\
	41.5	6.89305186257733\\
	42	6.97609451177925\\
	42.5	7.05913776484719\\
	43	7.14218155611871\\
	43.5	7.22522582707056\\
	44	7.30827052554262\\
	44.5	7.39131560504619\\
	45	7.47436102414739\\
};
\addlegendentry{\small $\frac{1}{2}\mathrm{log}_\mathrm{2}{(1+\SNR)}$};

\addplot [color=red,dashed,line width=1.0pt]
table[row sep=crcr]{%
	5	0.363812312929739\\
	5.5	0.395872400305428\\
	6	0.433735957778191\\
	6.5	0.477788167292683\\
	7	0.522436994906735\\
	7.5	0.567686862373678\\
	8	0.613569771560151\\
	8.5	0.661241349804746\\
	9	0.7203577620113\\
	9.5	0.780423073535524\\
	10	0.841432478801897\\
	10.5	0.903324286387758\\
	11	0.973359353623596\\
	11.5	1.04921690221651\\
	12	1.12545709448599\\
	12.5	1.20172858827388\\
	13	1.28750215852353\\
	13.5	1.37518994208931\\
	14	1.46154193618061\\
	14.5	1.55107992297435\\
	15	1.64536532640269\\
	15.5	1.73694964866861\\
	16	1.82909853047774\\
	16.5	1.92558655608641\\
	17	2.01885185865032\\
	17.5	2.11339775518312\\
	18	2.21061918336888\\
	18.5	2.30514688783299\\
	19	2.40471956313306\\
	19.5	2.50377230405928\\
	20	2.60488404875943\\
	20.5	2.70805863720317\\
	21	2.81246577417458\\
	21.5	2.91771969984885\\
	22	3.02451772969183\\
	22.5	3.12956847716218\\
	23	3.23348605288549\\
	23.5	3.3353591642491\\
	24	3.43275542707895\\
	24.5	3.52459483262552\\
	25	3.60964370335415\\
	25.5	3.68670008118292\\
	26	3.7553353253618\\
	26.5	3.81438177624735\\
	27	3.86378049296831\\
	27.5	3.90365361927849\\
	28	3.93468858088163\\
	28.5	3.95779185093405\\
	29	3.97412579798505\\
	29.5	3.98507058885478\\
	30	3.99194707297554\\
	30.5	3.99597275727091\\
	31	3.99815306552372\\
	31.5	3.99923029204145\\
	32	3.99971238564609\\
	32.5	3.99990480725961\\
	33	3.99997254113661\\
	33.5	3.99999320959073\\
	34	3.99999858775443\\
	34.5	3.99999975818047\\
	35	3.99999996672083\\
};
\addlegendentry{\small $\rateHDD$ uniform dist.};

\addplot [color=red,solid,line width=1.0pt]
table[row sep=crcr]{%
	16	2.36392144642074\\
	16.5	2.44504225527046\\
	17	2.52636074362662\\
	17.5	2.60785413242404\\
	18	2.68947476287496\\
	18.5	2.77116202003082\\
	19	2.85282606963498\\
	19.5	2.93433393002151\\
	20	3.01548454163904\\
	20.5	3.09602059725442\\
	21	3.17561254701753\\
	21.5	3.25387184148849\\
	22	3.33037072795361\\
	22.5	3.40463632250205\\
	23	3.47617106271669\\
	23.5	3.54447594306425\\
	24	3.6090676589179\\
	24.5	3.66944615505262\\
	25	3.72518334105358\\
	25.5	3.77588496435489\\
	26	3.82121255257153\\
	26.5	3.8609653592629\\
	27	3.89500842904543\\
	27.5	3.92337398073915\\
	28	3.94625626079628\\
	28.5	3.96400854672517\\
	29	3.97717105804268\\
	29.5	3.98638466013946\\
	30	3.99244296272019\\
	30.5	3.99613764039685\\
	31	3.99819743622726\\
	31.5	3.9992303171424\\
	32	3.99971225681673\\
	32.5	3.99990477523023\\
	33	3.99997253156726\\
	33.5	3.99999320339065\\
	34	3.99999858605238\\
	34.5	3.99999975783564\\
	35	3.99999996666426\\
};
\addlegendentry{\small $\rateHDD$ with shaping};

\end{axis}
\end{tikzpicture}%

%% file: differentASKorder_updatenew_georgrate.tikz
\definecolor{mycolor1}{rgb}{0.85098,0.32549,0.09804}%
\definecolor{mycolor2}{rgb}{0.00000,0.44706,0.74118}%
\definecolor{mycolor3}{rgb}{0.63529,0.07843,0.18431}%
\definecolor{mycolor4}{rgb}{0.49412,0.18431,0.55686}%
\definecolor{mycolor5}{rgb}{0.00000,0.49804,0.00000}%
\begin{tikzpicture}
\pgfplotsset{every tick label/.append style={font=\small}}
\tikzstyle{loosely dashdotted}=      [dash pattern=on 3pt off 4pt on \the\pgflinewidth off 4pt]
\begin{axis}[%
width=\figurewidth,
height=\figureheight,
at={(0\figurewidth,0\figureheight)},
scale only axis,
ylabel style={yshift=-0.65cm},
xlabel style={yshift=0.2cm},
xmin=5,
xmax=45,
ylabel={\small Spectral efficiency (bits/channel use)},
xlabel={\small $\SNR\, (\textrm{dB})$},
grid style={gray,opacity=0.15},
xmajorgrids,
ymin=0,
ymax=8,
xtick={ 5, 10, 15, 20, 25, 30, 35, 40, 45},
ymajorgrids,
axis background/.style={fill=white},
legend style={at={(0.02,0.98)},anchor=north west,legend cell align=left,align=left,draw=white!15!black}
]

\addplot [color=black,solid,line width=1.0pt,,mark size=1.5pt,mark=square,mark repeat={8} ,mark options={solid}]
table[row sep=crcr]{%
	0	0.5\\
	0.5	0.542718601409599\\
	1	0.587818317346194\\
	1.5	0.635275697101155\\
	2	0.685052334875493\\
	2.5	0.737095848832961\\
	3	0.791341177455778\\
	3.5	0.847712127039064\\
	4	0.906123095650313\\
	4.5	0.96648089622865\\
	5	1.0286866043034\\
	5.5	1.0926373631491\\
	6	1.15822808981313\\
	6.5	1.22535303798124\\
	7	1.29390718678102\\
	7.5	1.36378743725932\\
	8	1.43489360958514\\
	8.5	1.50712924350316\\
	9	1.58040221195651\\
	9.5	1.65462516310201\\
	10	1.72971580931865\\
	10.5	1.8055970835281\\
	11	1.88219718352143\\
	11.5	1.95944952434599\\
	12	2.03729261745271\\
	12.5	2.11566989350116\\
	13	2.19452948368152\\
	13.5	2.27382397230579\\
	14	2.35351013136442\\
	14.5	2.43354864582107\\
	15	2.51390383667526\\
	15.5	2.59454338728602\\
	16	2.6754380771243\\
	16.5	2.75656152600264\\
	17	2.83788995090244\\
	17.5	2.91940193676406\\
	18	3.00107822200099\\
	18.5	3.08290149902681\\
	19	3.16485622972095\\
	19.5	3.24692847548876\\
	20	3.3291057413759\\
	20.5	3.41137683356277\\
	21	3.49373172947796\\
	21.5	3.5761614597214\\
	22	3.65865800096827\\
	22.5	3.74121417902728\\
	23	3.8238235812452\\
	23.5	3.90648047747976\\
	24	3.98917974890062\\
	24.5	4.07191682392081\\
	25	4.1546876206064\\
	25.5	4.23748849495864\\
	26	4.3203161945087\\
	26.5	4.40316781671055\\
	27	4.48604077166048\\
	27.5	4.56893274871298\\
	28	4.65184168660144\\
	28.5	4.73476574670804\\
	29	4.81770328916082\\
	29.5	4.90065285146676\\
	30	4.983613129418\\
	30.5	5.0665829600343\\
	31	5.14956130632863\\
	31.5	5.23254724370396\\
	32	5.31553994780929\\
	32.5	5.39853868370022\\
	33	5.48154279616551\\
	33.5	5.5645517010955\\
	34	5.64756487778109\\
	34.5	5.73058186204385\\
	35	5.81360224010802\\
	35.5	5.89662564313503\\
	36	5.97965174234899\\
	36.5	6.06268024468985\\
	37	6.1457108889372\\
	37.5	6.22874344225396\\
	38	6.31177769710469\\
	38.5	6.394813468508\\
	39	6.47785059158696\\
	39.5	6.56088891938526\\
	40	6.64392832092027\\
	40.5	6.72696867944758\\
	41	6.81000989091377\\
	41.5	6.89305186257733\\
	42	6.97609451177925\\
	42.5	7.05913776484719\\
	43	7.14218155611871\\
	43.5	7.22522582707056\\
	44	7.30827052554262\\
	44.5	7.39131560504619\\
	45	7.47436102414739\\
};
\addlegendentry{\small $\frac{1}{2}\mathrm{log}_\mathrm{2}{(1+\SNR)}$};

\addplot [color=mycolor1,solid,dotted,line width=1.0pt]
  table[row sep=crcr]{%
5	0.783701935340367\\
5.5	0.847751704996402\\
6	0.913087068517602\\
6.5	0.979579047651123\\
7	1.04702054312745\\
7.5	1.11516035522601\\
8	1.18368061001903\\
8.5	1.2522124107264\\
9	1.32033674093093\\
9.5	1.38760721667943\\
10	1.45352729863252\\
10.5	1.51757343504965\\
11	1.57920749328205\\
11.5	1.63789496715365\\
12	1.69309441331653\\
12.5	1.74430346874887\\
13	1.79107047673952\\
13.5	1.83301901678481\\
14	1.86987091689484\\
14.5	1.90147910417417\\
15	1.92784459378852\\
15.5	1.94913569508031\\
16	1.96568801093199\\
16.5	1.97799735727286\\
17	1.98668843642428\\
17.5	1.99246481647231\\
18	1.99604509549245\\
18.5	1.99809384456866\\
19	1.9991649835125\\
19.5	1.99967120605256\\
20	1.99988501392206\\
20.5	1.99996476678914\\
21	1.99999070035476\\
21.5	1.99999792021753\\
22	1.99999961384282\\
22.5	1.99999994182415\\
23	1.99999999306923\\
23.5	1.99999999936563\\
24	1.99999999995681\\
24.5	1.99999999999789\\
25	1.99999999999993\\
  };
\addlegendentry{\small 4-ASK $\rateHDD$};

\addplot [color=mycolor2,solid,dashdotted,line width=1.0pt]
  table[row sep=crcr]{%
10.5	1.51702695762757\\
11	1.59361343827667\\
11.5	1.67082988987842\\
12	1.74858246084388\\
12.5	1.82675617774617\\
13	1.90519717593716\\
13.5	1.9837133886818\\
14	2.06204173053991\\
14.5	2.13986906190908\\
15	2.21682298988338\\
15.5	2.29248115748419\\
16	2.36639292358119\\
16.5	2.43806717993957\\
17	2.50700767471938\\
17.5	2.57271338485394\\
18	2.63468606494404\\
18.5	2.69245307521573\\
19	2.7455789443689\\
19.5	2.79368554835825\\
20	2.8364763167696\\
20.5	2.87375295328929\\
21	2.9054425671703\\
21.5	2.9316218740713\\
22	2.95252766994949\\
22.5	2.96856449131\\
23	2.98029432930195\\
23.5	2.98840359238926\\
24	2.99365229406782\\
24.5	2.9967991192209\\
25	2.99852811662064\\
25.5	2.99938911281046\\
26	2.999773767919\\
26.5	2.99992616839332\\
27	2.99997911727535\\
27.5	2.99999495055255\\
28	2.99999897624203\\
28.5	2.99999982970457\\
29	2.99999997731873\\
29.5	2.99999999764653\\
30	2.9999999998155\\
  };
\addlegendentry{\small 8-ASK $\rateHDD$};

\addplot [color=mycolor3,solid,dashed,line width=1.0pt]
  table[row sep=crcr]{%
16	2.36392144642074\\
16.5	2.44504225527046\\
17	2.52636074362662\\
17.5	2.60785413242404\\
18	2.68947476287496\\
18.5	2.77116202003082\\
19	2.85282606963498\\
19.5	2.93433393002151\\
20	3.01548454163904\\
20.5	3.09602059725442\\
21	3.17561254701753\\
21.5	3.25387184148849\\
22	3.33037072795361\\
22.5	3.40463632250205\\
23	3.47617106271669\\
23.5	3.54447594306425\\
24	3.6090676589179\\
24.5	3.66944615505262\\
25	3.72518334105358\\
25.5	3.77588496435489\\
26	3.82121255257153\\
26.5	3.8609653592629\\
27	3.89500842904543\\
27.5	3.92337398073915\\
28	3.94625626079628\\
28.5	3.96400854672517\\
29	3.97717105804268\\
29.5	3.98638466013946\\
30	3.99244296272019\\
30.5	3.99613764039685\\
31	3.99819743622726\\
31.5	3.9992303171424\\
32	3.99971225681673\\
32.5	3.99990477523023\\
33	3.99997253156726\\
33.5	3.99999320339065\\
34	3.99999858605238\\
34.5	3.99999975783564\\
35	3.99999996666426\\
  };
\addlegendentry{\small 16-ASK $\rateHDD$};

\addplot [color=mycolor4,solid,line width=1.0pt,loosely dashdotted]
  table[row sep=crcr]{%
22	3.33194853696669\\
22.5	3.41450319332293\\
23	3.49710704941473\\
23.5	3.57973767017895\\
24	3.66240486209448\\
24.5	3.74503633004095\\
25	3.82759371281117\\
25.5	3.90994246673573\\
26	3.99192251653509\\
26.5	4.07333104328904\\
27	4.15383673493956\\
27.5	4.23307162696957\\
28	4.31062941422118\\
28.5	4.386060232193\\
29	4.45883556060307\\
29.5	4.52845793774582\\
30	4.59443208103061\\
30.5	4.65614563468912\\
31	4.7133927435639\\
31.5	4.76558393191133\\
32	4.81236198222361\\
32.5	4.853417380381\\
33	4.88895762507801\\
33.5	4.91847169447067\\
34	4.94261489272943\\
34.5	4.96137841721494\\
35	4.97533329643228\\
35.5	4.98506313415783\\
36	4.99174286045105\\
36.5	4.99559907470469\\
37	4.9979479329504\\
37.5	4.99914003022537\\
38	4.99967632055237\\
38.5	4.99989203724161\\
39	4.99996856258948\\
39.5	4.99999214182018\\
40	4.99999834537412\\
  };
\addlegendentry{\small 32-ASK $\rateHDD$};

\addplot [color=mycolor5,solid,line width=1.0pt]
  table[row sep=crcr]{%
27.5	4.233\\
28	4.31380943957647\\
28.5	4.39672068895946\\
29	4.47963407935708\\
29.5	4.56251096137228\\
30	4.64559191253718\\
30.5	4.72838023538554\\
31	4.81112479389938\\
31.5	4.89385885844144\\
32	4.97608530586433\\
32.5	5.05779535062537\\
33	5.13815734947981\\
33.5	5.21826981091062\\
34	5.29650171865254\\
34.5	5.37241440953551\\
35	5.44523620194413\\
35.5	5.51626146750597\\
36	5.58388813259347\\
36.5	5.64588775447234\\
37	5.70418333249703\\
37.5	5.75789734478812\\
38	5.80482481624786\\
38.5	5.84774063250083\\
39	5.88410250936959\\
39.5	5.91299783864277\\
40	5.93990468436567\\
40.5	5.95904223324208\\
41	5.97193085605616\\
41.5	5.98254996498399\\
42	5.9905289652534\\
42.5	5.995230330111\\
43	5.99779197854429\\
43.5	5.99907022063\\
44	5.99964803908696\\
44.5	5.99988180365066\\
45	5.99996530505298\\
  };
\addlegendentry{\small 64-ASK $\rateHDD$};

\filldraw [draw=red,fill=red,solid] (axis cs:10.6,1.529) ellipse [x radius=3, y radius=6];
\filldraw [draw=red,fill=red,solid] (axis cs:16.29,2.43) ellipse [x radius=3, y radius=6];
\filldraw [draw=red,fill=red,solid] (axis cs:22.7,3.43) ellipse [x radius=3, y radius=6];
\filldraw [draw=red,fill=red,solid] (axis cs:28.4,4.39) ellipse [x radius=3, y radius=6];
\end{axis}

\end{tikzpicture}%

%% file: operating_point_update_new_map_newgeorg.tikz
\definecolor{mycolor1}{rgb}{0.00000,0.44706,0.74118}%
\definecolor{mycolor2}{rgb}{0.00000,0.49804,0.00000}%
\begin{tikzpicture}

\begin{axis}[%
%\pgfplotsset{every tick label/.append style={font=\small}}
width=\figurewidth,
height=\figureheight,
at={(0\figurewidth,0\figureheight)},
ylabel={\small Spectral efficiency (bits/channel use)},
xlabel={\small $\SNR\, (\textrm{dB})$},
ylabel style={yshift=-0.5cm},
xlabel style={yshift=0.2cm},
xmin=16,
xmajorgrids,
ymajorgrids,
xmax=32,
grid style={gray,opacity=0.15},
ymin=1.7,
ymax=4.5,
axis background/.style={fill=white},
legend style={at={(0.01,0.99)},anchor=north west,legend cell align=left,align=left,draw=white!15!black}
]

\addplot [color=mycolor2,solid,line width=1.0pt]
table[row sep=crcr]{%
	16	2.36392144642074\\
	16.5	2.44504225527046\\
	17	2.52636074362662\\
	17.5	2.60785413242404\\
	18	2.68947476287496\\
	18.5	2.77116202003082\\
	19	2.85282606963498\\
	19.5	2.93433393002151\\
	20	3.01548454163904\\
	20.5	3.09602059725442\\
	21	3.17561254701753\\
	21.5	3.25387184148849\\
	22	3.33037072795361\\
	22.5	3.40463632250205\\
	23	3.47617106271669\\
	23.5	3.54447594306425\\
	24	3.6090676589179\\
	24.5	3.66944615505262\\
	25	3.72518334105358\\
	25.5	3.77588496435489\\
	26	3.82121255257153\\
	26.5	3.8609653592629\\
	27	3.89500842904543\\
	27.5	3.92337398073915\\
	28	3.94625626079628\\
	28.5	3.96400854672517\\
	29	3.97717105804268\\
	29.5	3.98638466013946\\
	30	3.99244296272019\\
	30.5	3.99613764039685\\
	31	3.99819743622726\\
	31.5	3.9992303171424\\
	32	3.99971225681673\\
	32.5	3.99990477523023\\
	33	3.99997253156726\\
	33.5	3.99999320339065\\
	34	3.99999858605238\\
	34.5	3.99999975783564\\
	35	3.99999996666426\\
};
\addlegendentry{\small $\rateHDD$};

\addplot [color=mycolor1,solid,line width=1.0pt,mark size=1.5pt,mark=*,mark repeat={3} ,mark options={solid}]
  table[row sep=crcr]{%
5	0.141660470080237\\
5.5	0.192532428464601\\
6	0.249930399993503\\
6.5	0.310812773316696\\
7	0.374191901131067\\
7.5	0.442509203192164\\
8	0.510426911192567\\
8.5	0.582230812899372\\
9	0.655203611156059\\
9.5	0.729390522608177\\
10	0.804410626099275\\
10.5	0.880390368650076\\
11	0.956641050924026\\
11.5	1.03394362571122\\
12	1.11227833186875\\
12.5	1.1902476800928\\
13	1.26899124939578\\
13.5	1.34857746368699\\
14	1.42819135951011\\
14.5	1.5077905817247\\
15	1.58866564698154\\
15.5	1.66875185366017\\
16	1.7496178305035\\
16.5	1.83206681833528\\
17	1.91159618423536\\
17.5	1.99437922489654\\
18	2.07551302007168\\
18.5	2.15560952597801\\
19	2.23579202152987\\
19.5	2.31462424107039\\
20	2.39027927734412\\
20.5	2.46068010085386\\
21	2.5313329554367\\
21.5	2.59720297710518\\
22	2.65678323341968\\
22.5	2.70887256060072\\
23	2.75665588588401\\
23.5	2.80337118209493\\
24	2.84080827492591\\
24.5	2.87277567866\\
25	2.90255573744699\\
25.5	2.92671617845842\\
26	2.94570976356006\\
26.5	2.96237396227717\\
27	2.97456148245287\\
27.5	2.98459997286548\\
28	2.98999880106067\\
28.5	2.99518909824628\\
29	2.99742944605562\\
29.5	2.99898595099643\\
30	2.99963260468839\\
30.5	2.99983619996687\\
31	2.99995892276039\\
31.5	3\\
32	3\\
32.5	3\\
33	3\\
33.5	3\\
34	3\\
34.5	3\\
35	3\\
  };
\addlegendentry{\small $\ent(A)$};

%\addplot [color=mycolor1,dotted,line width=1.0pt]
%table[row sep=crcr]{%
%5	0.443960470080237\\
%5.5	0.494832428464601\\
%6	0.552230399993503\\
%6.5	0.613112773316696\\
%7	0.676491901131067\\
%7.5	0.744809203192164\\
%8	0.812726911192567\\
%8.5	0.884530812899372\\
%9	0.957503611156059\\
%9.5	1.03169052260818\\
%10	1.10671062609927\\
%10.5	1.18269036865008\\
%11	1.25894105092403\\
%11.5	1.33624362571122\\
%12	1.41457833186875\\
%12.5	1.4925476800928\\
%13	1.57129124939578\\
%13.5	1.65087746368699\\
%14	1.73049135951011\\
%14.5	1.8100905817247\\
%15	1.89096564698154\\
%15.5	1.97105185366017\\
%16	2.0519178305035\\
%16.5	2.13436681833528\\
%17	2.21389618423536\\
%17.5	2.29667922489654\\
%18	2.37781302007168\\
%18.5	2.45790952597801\\
%19	2.53809202152987\\
%19.5	2.61692424107039\\
%20	2.69257927734412\\
%20.5	2.76298010085386\\
%21	2.8336329554367\\
%21.5	2.89950297710518\\
%22	2.95908323341968\\
%22.5	3.01117256060072\\
%23	3.05895588588401\\
%23.5	3.10567118209493\\
%24	3.14310827492591\\
%24.5	3.17507567866\\
%25	3.20485573744699\\
%25.5	3.22901617845842\\
%26	3.24800976356006\\
%26.5	3.26467396227717\\
%27	3.27686148245287\\
%27.5	3.28689997286548\\
%28	3.29229880106067\\
%28.5	3.29748909824628\\
%29	3.29972944605562\\
%29.5	3.30128595099643\\
%30	3.30193260468839\\
%30.5	3.30213619996687\\
%31	3.30225892276039\\
%31.5	3.3023\\
%32	3.3023\\
%32.5	3.3023\\
%33	3.3023\\
%33.5	3.3023\\
%34	3.3023\\
%34.5	3.3023\\
%35	3.3023\\
%};
%\addlegendentry{\small $\ent(A)+0.3023$};

\addplot [color=mycolor1,dotted,line width=1.0pt]
table[x expr=\thisrow{X}*1, y expr=\thisrow{Y}*1+0.3617]{
	X Y
	5	0.141660470080237
	5.5	0.192532428464601
	6	0.249930399993503
	6.5	0.310812773316696
	7	0.374191901131067
	7.5	0.442509203192164
	8	0.510426911192567
	8.5	0.582230812899372
	9	0.655203611156059
	9.5	0.729390522608177
	10	0.804410626099275
	10.5	0.880390368650076
	11	0.956641050924026
	11.5	1.03394362571122
	12	1.11227833186875
	12.5	1.1902476800928
	13	1.26899124939578
	13.5	1.34857746368699
	14	1.42819135951011
	14.5	1.5077905817247
	15	1.58866564698154
	15.5	1.66875185366017
	16	1.7496178305035
	16.5	1.83206681833528
	17	1.91159618423536
	17.5	1.99437922489654
	18	2.07551302007168
	18.5	2.15560952597801
	19	2.23579202152987
	19.5	2.31462424107039
	20	2.39027927734412
	20.5	2.46068010085386
	21	2.5313329554367
	21.5	2.59720297710518
	22	2.65678323341968
	22.5	2.70887256060072
	23	2.75665588588401
	23.5	2.80337118209493
	24	2.84080827492591
	24.5	2.87277567866
	25	2.90255573744699
	25.5	2.92671617845842
	26	2.94570976356006
	26.5	2.96237396227717
	27	2.97456148245287
	27.5	2.98459997286548
	28	2.98999880106067
	28.5	2.99518909824628
	29	2.99742944605562
	29.5	2.99898595099643
	30	2.99963260468839
	30.5	2.99983619996687
	31	2.99995892276039
	31.5	3
	32	3
	32.5	3
	33	3
	33.5	3
	34	3
	34.5	3
	35	3
};
\addlegendentry{\small $\ent(A)+0.3617$};

\addplot [color=mycolor1,dashdotted,line width=1.0pt]
  table[row sep=crcr]{%
5	0.822560470080237\\
5.5	0.873432428464601\\
6	0.930830399993503\\
6.5	0.991712773316696\\
7	1.05509190113107\\
7.5	1.12340920319216\\
8	1.19132691119257\\
8.5	1.26313081289937\\
9	1.33610361115606\\
9.5	1.41029052260818\\
10	1.48531062609927\\
10.5	1.56129036865008\\
11	1.63754105092403\\
11.5	1.71484362571122\\
12	1.79317833186875\\
12.5	1.8711476800928\\
13	1.94989124939578\\
13.5	2.02947746368699\\
14	2.10909135951011\\
14.5	2.1886905817247\\
15	2.26956564698154\\
15.5	2.34965185366017\\
16	2.4305178305035\\
16.5	2.51296681833528\\
17	2.59249618423536\\
17.5	2.67527922489654\\
18	2.75641302007168\\
18.5	2.83650952597801\\
19	2.91669202152987\\
19.5	2.99552424107039\\
20	3.07117927734412\\
20.5	3.14158010085386\\
21	3.2122329554367\\
21.5	3.27810297710518\\
22	3.33768323341968\\
22.5	3.38977256060072\\
23	3.43755588588401\\
23.5	3.48427118209493\\
24	3.52170827492591\\
24.5	3.55367567866\\
25	3.58345573744699\\
25.5	3.60761617845842\\
26	3.62660976356006\\
26.5	3.64327396227717\\
27	3.65546148245287\\
27.5	3.66549997286547\\
28	3.67089880106067\\
28.5	3.67608909824628\\
29	3.67832944605562\\
29.5	3.67988595099643\\
30	3.68053260468839\\
30.5	3.68073619996687\\
31	3.68085892276039\\
31.5	3.6809\\
32	3.6809\\
32.5	3.6809\\
33	3.6809\\
33.5	3.6809\\
34	3.6809\\
34.5	3.6809\\
35	3.6809\\
 };
\addlegendentry{\small $\ent(A)+0.5946$};

\addplot [color=mycolor1,dashed,line width=1.0pt]
  table[row sep=crcr]{%
5	0.891660470080237\\
5.5	0.942532428464601\\
6	0.999930399993503\\
6.5	1.0608127733167\\
7	1.12419190113107\\
7.5	1.19250920319216\\
8	1.26042691119257\\
8.5	1.33223081289937\\
9	1.40520361115606\\
9.5	1.47939052260818\\
10	1.55441062609927\\
10.5	1.63039036865008\\
11	1.70664105092403\\
11.5	1.78394362571122\\
12	1.86227833186875\\
12.5	1.9402476800928\\
13	2.01899124939578\\
13.5	2.09857746368699\\
14	2.17819135951011\\
14.5	2.2577905817247\\
15	2.33866564698154\\
15.5	2.41875185366017\\
16	2.4996178305035\\
16.5	2.58206681833528\\
17	2.66159618423536\\
17.5	2.74437922489654\\
18	2.82551302007168\\
18.5	2.90560952597801\\
19	2.98579202152987\\
19.5	3.06462424107039\\
20	3.14027927734412\\
20.5	3.21068010085386\\
21	3.2813329554367\\
21.5	3.34720297710518\\
22	3.40678323341968\\
22.5	3.45887256060072\\
23	3.50665588588401\\
23.5	3.55337118209493\\
24	3.59080827492591\\
24.5	3.62277567866\\
25	3.65255573744699\\
25.5	3.67671617845842\\
26	3.69570976356006\\
26.5	3.71237396227717\\
27	3.72456148245287\\
27.5	3.73459997286548\\
28	3.73999880106067\\
28.5	3.74518909824628\\
29	3.74742944605562\\
29.5	3.74898595099643\\
30	3.74963260468839\\
30.5	3.74983619996687\\
31	3.74995892276039\\
31.5	3.75\\
32	3.75\\
32.5	3.75\\
33	3.75\\
33.5	3.75\\
34	3.75\\
34.5	3.75\\
35	3.75\\
  };
\addlegendentry{\small $\ent(A)+0.75$};

\addplot [color=red,solid,line width=1.0pt,mark size=3.0pt,only marks,mark=+,mark options={solid},opacity=1]
table[row sep=crcr]{%
22.1648 3.3549\\
23.66 3.5661\\
};

\draw [dashed,red,->] (axis cs:22.1648,3.3549) to [bend right=45] (axis cs:24,2.2); 
\draw [dashed,red,->] (axis cs:23.66,3.5661) to [bend right=45] (axis cs:27,2.75); 
\node[red] at (axis cs:27.2,2.1) {\small Crossing point for $\gamma=0.5946$};
\node[red] at (axis cs:27.2,2.65) {\small Crossing point for $\gamma=0.75$};

\end{axis}

\end{tikzpicture}%

%% file: F1_16ASK_illustration_new_MAP_QAMupdate_georgrate.tikz
\definecolor{mycolor1}{rgb}{0.49400,0.18400,0.55600}%
\definecolor{mycolor2}{rgb}{0.00000,0.44706,0.74118}%
\definecolor{mycolor3}{rgb}{0.00000,0.49804,0.00000}%
\begin{tikzpicture}

\begin{axis}[%
%\pgfplotsset{every tick label/.append style={font=\small}}
width=\figurewidth,
height=\figureheight,
at={(0\figurewidth,0\figureheight)},
ylabel={\small Spectral efficiency (bits/channel use)},
xlabel={\small $\SNR\, (\textrm{dB})$},
ylabel style={yshift=-0.5cm},
xlabel style={yshift=0.2cm},
xmin=18,
xmax=28,
ymin=3.5,
ymax=8,
ytick={3.5,4,4.5,5,5.5,6,6.5,7,7.5,8},
grid style={gray,opacity=0.15},
xmajorgrids,
ymajorgrids,
axis background/.style={fill=white},
legend style={at={(0.98,0.02)},anchor=south east,legend cell align=left,align=left,draw=white!15!black}
]
\addplot [color=mycolor1,solid,line width=1.0pt]
  table[x expr=\thisrow{X}*1, y expr=\thisrow{Y}*2]{
X Y
5	0.363812312929739
5.5	0.395872400305428
6	0.433735957778191
6.5	0.477788167292683
7	0.522436994906735
7.5	0.567686862373678
8	0.613569771560151
8.5	0.661241349804746
9	0.7203577620113
9.5	0.780423073535524
10	0.841432478801897
10.5	0.903324286387758
11	0.973359353623596
11.5	1.04921690221651
12	1.12545709448599
12.5	1.20172858827388
13	1.28750215852353
13.5	1.37518994208931
14	1.46154193618061
14.5	1.55107992297435
15	1.64536532640269
15.5	1.73694964866861
16	1.82909853047774
16.5	1.92558655608641
17	2.01885185865032
17.5	2.11339775518312
18	2.21061918336888
18.5	2.30514688783299
19	2.40471956313306
19.5	2.50377230405928
20	2.60488404875943
20.5	2.70805863720317
21	2.81246577417458
21.5	2.91771969984885
22	3.02451772969183
22.5	3.12956847716218
23	3.23348605288549
23.5	3.3353591642491
24	3.43275542707895
24.5	3.52459483262552
25	3.60964370335415
25.5	3.68670008118292
26	3.7553353253618
26.5	3.81438177624735
27	3.86378049296831
27.5	3.90365361927849
28	3.93468858088163
28.5	3.95779185093405
29	3.97412579798505
29.5	3.98507058885478
30	3.99194707297554
30.5	3.99597275727091
31	3.99815306552372
31.5	3.99923029204145
32	3.99971238564609
32.5	3.99990480725961
33	3.99997254113661
33.5	3.99999320959073
34	3.99999858775443
34.5	3.99999975818047
35	3.99999996672083
};
\addlegendentry{\small $\rateHDD$ (uniform)};

\addplot [color=mycolor3,solid,line width=1.0pt,mark size=1.5pt,mark=*,mark repeat={5} ,mark options={solid}]
table[x expr=\thisrow{X}*1, y expr=\thisrow{Y}*2]{
	X Y
	16	2.36392144642074
	16.5	2.44504225527046
	17	2.52636074362662
	17.5	2.60785413242404
	18	2.68947476287496
	18.5	2.77116202003082
	19	2.85282606963498
	19.5	2.93433393002151
	20	3.01548454163904
	20.5	3.09602059725442
	21	3.17561254701753
	21.5	3.25387184148849
	22	3.33037072795361
	22.5	3.40463632250205
	23	3.47617106271669
	23.5	3.54447594306425
	24	3.6090676589179
	24.5	3.66944615505262
	25	3.72518334105358
	25.5	3.77588496435489
	26	3.82121255257153
	26.5	3.8609653592629
	27	3.89500842904543
	27.5	3.92337398073915
	28	3.94625626079628
	28.5	3.96400854672517
	29	3.97717105804268
	29.5	3.98638466013946
	30	3.99244296272019
	30.5	3.99613764039685
	31	3.99819743622726
	31.5	3.9992303171424
	32	3.99971225681673
	32.5	3.99990477523023
	33	3.99997253156726
	33.5	3.99999320339065
	34	3.99999858605238
	34.5	3.99999975783564
	35	3.99999996666426
};
\addlegendentry{\small $\rateHDD$ (shaping)};

\addplot [color=mycolor2,solid,line width=1.0pt,mark size=1.5pt,mark=square,mark repeat={5} ,mark options={solid}]
table[x expr=\thisrow{X}*1, y expr=\thisrow{Y}*2]{
	X Y
	5	0.141660470080237
	5.5	0.192532428464601
	6	0.249930399993503
	6.5	0.310812773316696
	7	0.374191901131067
	7.5	0.442509203192164
	8	0.510426911192567
	8.5	0.582230812899372
	9	0.655203611156059
	9.5	0.729390522608177
	10	0.804410626099275
	10.5	0.880390368650076
	11	0.956641050924026
	11.5	1.03394362571122
	12	1.11227833186875
	12.5	1.1902476800928
	13	1.26899124939578
	13.5	1.34857746368699
	14	1.42819135951011
	14.5	1.5077905817247
	15	1.58866564698154
	15.5	1.66875185366017
	16	1.7496178305035
	16.5	1.83206681833528
	17	1.91159618423536
	17.5	1.99437922489654
	18	2.07551302007168
	18.5	2.15560952597801
	19	2.23579202152987
	19.5	2.31462424107039
	20	2.39027927734412
	20.5	2.46068010085386
	21	2.5313329554367
	21.5	2.59720297710518
	22	2.65678323341968
	22.5	2.70887256060072
	23	2.75665588588401
	23.5	2.80337118209493
	24	2.84080827492591
	24.5	2.87277567866
	25	2.90255573744699
	25.5	2.92671617845842
	26	2.94570976356006
	26.5	2.96237396227717
	27	2.97456148245287
	27.5	2.98459997286548
	28	2.98999880106067
	28.5	2.99518909824628
	29	2.99742944605562
	29.5	2.99898595099643
	30	2.99963260468839
	30.5	2.99983619996687
	31	2.99995892276039
	31.5	3
	32	3
	32.5	3
	33	3
	33.5	3
	34	3
	34.5	3
	35	3
};
\addlegendentry{\small $2\ent(A)$};

\addplot [color=mycolor2,dashed,line width=1.0pt]
table[x expr=\thisrow{X}*1, y expr=\thisrow{Y}*2+0.7234]{
	X Y
	5	0.141660470080237
	5.5	0.192532428464601
	6	0.249930399993503
	6.5	0.310812773316696
	7	0.374191901131067
	7.5	0.442509203192164
	8	0.510426911192567
	8.5	0.582230812899372
	9	0.655203611156059
	9.5	0.729390522608177
	10	0.804410626099275
	10.5	0.880390368650076
	11	0.956641050924026
	11.5	1.03394362571122
	12	1.11227833186875
	12.5	1.1902476800928
	13	1.26899124939578
	13.5	1.34857746368699
	14	1.42819135951011
	14.5	1.5077905817247
	15	1.58866564698154
	15.5	1.66875185366017
	16	1.7496178305035
	16.5	1.83206681833528
	17	1.91159618423536
	17.5	1.99437922489654
	18	2.07551302007168
	18.5	2.15560952597801
	19	2.23579202152987
	19.5	2.31462424107039
	20	2.39027927734412
	20.5	2.46068010085386
	21	2.5313329554367
	21.5	2.59720297710518
	22	2.65678323341968
	22.5	2.70887256060072
	23	2.75665588588401
	23.5	2.80337118209493
	24	2.84080827492591
	24.5	2.87277567866
	25	2.90255573744699
	25.5	2.92671617845842
	26	2.94570976356006
	26.5	2.96237396227717
	27	2.97456148245287
	27.5	2.98459997286548
	28	2.98999880106067
	28.5	2.99518909824628
	29	2.99742944605562
	29.5	2.99898595099643
	30	2.99963260468839
	30.5	2.99983619996687
	31	2.99995892276039
	31.5	3
	32	3
	32.5	3
	33	3
	33.5	3
	34	3
	34.5	3
	35	3
};
\addlegendentry{\small $2\ent(A)+0.7234$};

\addplot [color=mycolor2,dashdotted,line width=1.0pt]
  table[x expr=\thisrow{X}*1, y expr=\thisrow{Y}*2]{
  	X Y
5	0.891660470080237
5.5	0.942532428464601
6	0.999930399993503
6.5	1.0608127733167
7	1.12419190113107
7.5	1.19250920319216
8	1.26042691119257
8.5	1.33223081289937
9	1.40520361115606
9.5	1.47939052260818
10	1.55441062609927
10.5	1.63039036865008
11	1.70664105092403
11.5	1.78394362571122
12	1.86227833186875
12.5	1.9402476800928
13	2.01899124939578
13.5	2.09857746368699
14	2.17819135951011
14.5	2.2577905817247
15	2.33866564698154
15.5	2.41875185366017
16	2.4996178305035
16.5	2.58206681833528
17	2.66159618423536
17.5	2.74437922489654
18	2.82551302007168
18.5	2.90560952597801
19	2.98579202152987
19.5	3.06462424107039
20	3.14027927734412
20.5	3.21068010085386
21	3.2813329554367
21.5	3.34720297710518
22	3.40678323341968
22.5	3.45887256060072
23	3.50665588588401
23.5	3.55337118209493
24	3.59080827492591
24.5	3.62277567866
25	3.65255573744699
25.5	3.67671617845842
26	3.69570976356006
26.5	3.71237396227717
27	3.72456148245287
27.5	3.73459997286548
28	3.73999880106067
28.5	3.74518909824628
29	3.74742944605562
29.5	3.74898595099643
30	3.74963260468839
30.5	3.74983619996687
31	3.74995892276039
31.5	3.75
32	3.75
32.5	3.75
33	3.75
33.5	3.75
34	3.75
34.5	3.75
35	3.75
  };
\addlegendentry{\small $2\ent(A)+1.5$};

\addplot [color=mycolor3,line width=1.5pt,mark size=3pt,only marks,mark=x,mark options={solid}]
table[x expr=\thisrow{X}*1, y expr=\thisrow{Y}*2]{
	X Y  	
	25.02	3.6516
	24.38	3.5552
	23.342	3.3804
	22.49	3.2297
	21.236	3.0031
	20.35	2.8386
	18.85	2.5703

};
\addlegendentry{\small Simulation (shaping)};

\addplot [color=red,solid,line width=1.0pt,mark size=3.0pt,only marks,mark=+,mark options={solid},opacity=1]
table[x expr=\thisrow{X}*1, y expr=\thisrow{Y}*2]{
	X Y
	23.66 3.5661
};	

\draw[<->,black]  (axis cs:23.56,7.5322) to node[above] {\small $1.36$ dB} (axis cs:25.12,7.5322);

%\addplot [color=mycolor3,line width=1.1pt,mark size=2.0pt,only marks,mark=square,mark options={solid},opacity=0.7,forget plot]
%table[x expr=\thisrow{X}*1, y expr=\thisrow{Y}*2]{
%	X Y
%	17.25 2.267
%	18.55 2.4661      
%};

\addplot [color=mycolor3,line width=1.1pt,mark size=2.0pt,only marks,mark=square,mark options={solid},opacity=0.7,forget plot]
table[x expr=\thisrow{X}*1, y expr=\thisrow{Y}*2]{
	X Y
%	19 2.5927
	20.2 2.7749
	19.4 2.6564
	19.8 2.7173      
};

\end{axis}
\end{tikzpicture}%

%% file: F2_16ASK_illustration_new_MAP_QAMupdate_georgrate.tikz
\definecolor{mycolor1}{rgb}{0.00000,0.49804,0.00000}%
\definecolor{mycolor2}{rgb}{0.00000,0.44706,0.74118}%
\definecolor{mycolor3}{rgb}{0.49412,0.18431,0.55686}%
\begin{tikzpicture}

\begin{axis}[%
%\pgfplotsset{every tick label/.append style={font=\small}}
width=\figurewidth,
height=\figureheight,
at={(0\figurewidth,0\figureheight)},
ylabel={\small Spectral efficiency (bits/channel use)},
xlabel={\small $\SNR\, (\textrm{dB})$},
ylabel style={yshift=-0.5cm},
xlabel style={yshift=0.2cm},
xmin=16,
xmax=26,
xtick={12, 14, 16, 18, 20, 22, 24, 25, 27, 29, 31, 33},
grid style={gray,opacity=0.15},
xmajorgrids,
ymin=4.4,
ymax=8,
ytick={  4, 4.4, 4.8, 5.2, 5.6,   6, 6.4, 6.8, 7.2, 7.6,   8},
ymajorgrids,
axis background/.style={fill=white},
legend style={at={(0.01,0.99)},anchor=north west,legend cell align=left,align=left,draw=white!15!black}
]
\addplot [color=mycolor1,solid,line width=1.0pt,mark size=1.5pt,mark=*,mark repeat={4} ,mark options={solid}]
  table[x expr=\thisrow{X}*1, y expr=\thisrow{Y}*2]{
  	X Y
16	2.36392144642074
16.5	2.44504225527046
17	2.52636074362662
17.5	2.60785413242404
18	2.68947476287496
18.5	2.77116202003082
19	2.85282606963498
19.5	2.93433393002151
20	3.01548454163904
20.5	3.09602059725442
21	3.17561254701753
21.5	3.25387184148849
22	3.33037072795361
22.5	3.40463632250205
23	3.47617106271669
23.5	3.54447594306425
24	3.6090676589179
24.5	3.66944615505262
25	3.72518334105358
25.5	3.77588496435489
26	3.82121255257153
26.5	3.8609653592629
27	3.89500842904543
27.5	3.92337398073915
28	3.94625626079628
28.5	3.96400854672517
29	3.97717105804268
29.5	3.98638466013946
30	3.99244296272019
30.5	3.99613764039685
31	3.99819743622726
31.5	3.9992303171424
32	3.99971225681673
32.5	3.99990477523023
33	3.99997253156726
33.5	3.99999320339065
34	3.99999858605238
34.5	3.99999975783564
35	3.99999996666426
  };
\addlegendentry{\small $\rateHDD$ $256$-QAM (shaping)};

\addplot [color=mycolor2,solid,line width=1.0pt]
  table[x expr=\thisrow{X}*1, y expr=\thisrow{Y}*2]{
  	X Y
10.5	1.51702695762757
11	1.59361343827667
11.5	1.67082988987842
12	1.74858246084388
12.5	1.82675617774617
13	1.90519717593716
13.5	1.9837133886818
14	2.06204173053991
14.5	2.13986906190908
15	2.21682298988338
15.5	2.29248115748419
16	2.36639292358119
16.5	2.43806717993957
17	2.50700767471938
17.5	2.57271338485394
18	2.63468606494404
18.5	2.69245307521573
19	2.7455789443689
19.5	2.79368554835825
20	2.8364763167696
20.5	2.87375295328929
21	2.9054425671703
21.5	2.9316218740713
22	2.95252766994949
22.5	2.96856449131
23	2.98029432930195
23.5	2.98840359238926
24	2.99365229406782
24.5	2.9967991192209
25	2.99852811662064
25.5	2.99938911281046
26	2.999773767919
26.5	2.99992616839332
27	2.99997911727535
27.5	2.99999495055255
28	2.99999897624203
28.5	2.99999982970457
29	2.99999997731873
29.5	2.99999999764653
30	2.9999999998155
  };
\addlegendentry{\small $\rateHDD$ $64$-QAM (shaping)};

\addplot [color=mycolor3,dotted,line width=1.2pt]
  table[x expr=\thisrow{X}*1, y expr=\thisrow{Y}*2]{
X Y  	
5	0.508487824875371
5.5	0.559647626294144
6	0.614376157817358
6.5	0.672432703673546
7	0.73365355168849
7.5	0.797811272770864
8	0.864618099539084
8.5	0.933712153056884
9	1.00476588365755
9.5	1.07744260600083
10	1.1514683347023
10.5	1.22665130760315
11	1.30302331043247
11.5	1.38048602377436
12	1.45929182025147
12.5	1.53956656927841
13	1.62154544737791
13.5	1.70540988259773
14	1.79131742669857
14.5	1.87917560865191
15	1.96878219042002
15.5	2.0597406695818
16	2.15139996557569
16.5	2.24296277735994
17	2.33341703075116
17.5	2.42166170137155
18	2.5064793269472
18.5	2.58665130342344
19	2.66100573677511
19.5	2.72848548384597
20	2.78825040466836
20.5	2.83973987389702
21	2.88274050290557
21.5	2.91740732718518
22	2.94426519898137
22.5	2.96416184266275
23	2.97817461521429
23.5	2.98749747574809
24	2.99331476516894
24.5	2.99669198555361
25	2.99849987363992
25.5	2.99938330813838
26	2.99977297275004
26.5	2.99992617683923
27	2.99997912028541
27.5	2.99999495147138
28	2.99999897647697
28.5	2.99999982975382
29	2.999999977327
29.5	2.99999999764762
30	2.99999999981561
30.5	2.99999999998945
31	2.99999999999958
31.5	2.99999999999999
32	3
32.5	3
33	3
33.5	3
34	3
34.5	3
35	3
35.5	3
36	3
36.5	3
37	3
37.5	3
38	3
38.5	3
39	3
39.5	3
40	3
};
\addlegendentry{\small $\rateHDD$ $64$-QAM (uniform)};

\addplot [color=mycolor1,line width=1.5pt,mark size=3pt,only marks,mark=x,mark options={solid}]
  table[x expr=\thisrow{X}*1, y expr=\thisrow{Y}*2]{
X Y  	
	25.02	3.6516
	24.38	3.5552
	23.342	3.3804
	22.49	3.2297
	21.236	3.0031
	20.35	2.8386
	18.85	2.5703
};
\addlegendentry{\small Simulation $256$-QAM (shaping)};

\addplot [color=mycolor2,line width=1pt,mark size=2.4pt,only marks,mark=o,mark options={solid}]
  table[x expr=\thisrow{X}*1, y expr=\thisrow{Y}*2]{
X Y  	
19.746	2.7618
19.05	2.6724
17.63	2.4426
16.727	2.2702
};
\addlegendentry{\small Simulation $64$-QAM (shaping)};

\end{axis}
\end{tikzpicture}%

%% file: JLT_PAS_HDD_v29_arxiv.bbl
\begin{thebibliography}{10}
	\providecommand{\url}[1]{#1}
	\csname url@samestyle\endcsname
	\providecommand{\newblock}{\relax}
	\providecommand{\bibinfo}[2]{#2}
	\providecommand{\BIBentrySTDinterwordspacing}{\spaceskip=0pt\relax}
	\providecommand{\BIBentryALTinterwordstretchfactor}{4}
	\providecommand{\BIBentryALTinterwordspacing}{\spaceskip=\fontdimen2\font plus
		\BIBentryALTinterwordstretchfactor\fontdimen3\font minus
		\fontdimen4\font\relax}
	\providecommand{\BIBforeignlanguage}[2]{{%
			\expandafter\ifx\csname l@#1\endcsname\relax
			\typeout{** WARNING: IEEEtran.bst: No hyphenation pattern has been}%
			\typeout{** loaded for the language `#1'. Using the pattern for}%
			\typeout{** the default language instead.}%
			\else
			\language=\csname l@#1\endcsname
			\fi
			#2}}
	\providecommand{\BIBdecl}{\relax}
	\BIBdecl
	
	\bibitem{She17c}
	A.~Sheikh, A.~{Graell i Amat}, and G.~Liva, ``Probabilistically-shaped coded
	modulation with hard decision decoding for coherent optical systems,'' in
	\emph{Proc.\ European Conf. Optical Communications (ECOC)}, Gothenburg,
	Sweden, Sep. 2017.
	
	\bibitem{Barsoum_2007}
	M.~F. Barsoum, C.~Jones, and M.~Fitz, ``Constellation design via capacity
	maximization,'' in \emph{Proc.\ IEEE Int.\ Symp.\ Inf.\ Theory (ISIT)}, Nice,
	2007, pp. 1821--1825.
	
	\bibitem{polar_modulation}
	Z.~H. Peric, I.~B. Djordjevic, S.~M. Bogosavljevic, and M.~C. Stefanovic,
	``Design of signal constellations for {Gaussian} channel by using iterative
	polar quantization,'' in \emph{Proc. Mediterranean Electrotechnical Conf.,
		(MELECON)}, vol.~2, Tel-Aviv, 1998, pp. 866--869.
	
	\bibitem{Djordjevic_2010}
	I.~B. Djordjevic, H.~G. Batshon, L.~Xu, and T.~Wang, ``Coded
	polarization-multiplexed iterative polar modulation ({PM-IPM}) for beyond 400
	{G}b/s serial optical transmission,'' in \emph{Proc.\ Optical Fiber Commun.
		Conf. (OFC)}, San Diego, CA, 2010, pp. 1--3.
	
	\bibitem{Liu_2014}
	T.~Liu and I.~B. Djordjevic, ``Multidimensional optimal signal constellation
	sets and symbol mappings for block-interleaved coded-modulation enabling
	ultrahigh-speed optical transport,'' \emph{IEEE Photon.\ J.}, vol.~6, no.~4,
	pp. 1--14, Aug. 2014.
	
	\bibitem{Geller_2016}
	O.~Geller, R.~Dar, M.~Feder, and M.~Shtaif, ``A shaping algorithm for
	mitigating inter-channel nonlinear phase-noise in nonlinear fiber systems,''
	\emph{IEEE/OSA J.\ Lightw.\ Technol.}, vol.~34, no.~16, pp. 3884--3889, Aug.
	2016.
	
	\bibitem{Calderbank_1990}
	A.~R. Calderbank and L.~H. Ozarow, ``Nonequiprobable signaling on the gaussian
	channel,'' \emph{IEEE Trans.\ Inf.\ Theory}, vol.~36, no.~4, pp. 726--740,
	Jul. 1990.
	
	\bibitem{Forney_1992}
	G.~D. Forney, ``Trellis shaping,'' \emph{IEEE Trans.\ Inf.\ Theory}, vol.~38,
	no.~2, pp. 281--300, Mar. 1992.
	
	\bibitem{georg_tcom}
	G.~B\"ocherer, F.~Steiner, and P.~Schulte, ``Bandwidth efficient and
	rate-matched low-density parity-check coded modulation,'' \emph{IEEE Trans.\
		Commun.}, vol.~63, no.~12, pp. 4651--4665, Dec. 2015.
	
	\bibitem{Smith_2012}
	B.~P. Smith and F.~R. Kschischang, ``A pragmatic coded modulation scheme for
	high-spectral-efficiency fiber-optic communications,'' \emph{IEEE/OSA J.\
		Lightw.\ Technol.}, vol.~30, no.~13, pp. 2047--2053, Jul. 2012.
	
	\bibitem{Buchali1_2015}
	F.~Buchali, G.~Böcherer, W.~Idler, L.~Schmalen, P.~Schulte, and F.~Steiner,
	``Experimental demonstration of capacity increase and rate-adaptation by
	probabilistically shaped 64-{QAM},'' in \emph{Proc.\ European Conf. Optical
		Communications (ECOC)}, Valencia, 2015, pp. 1--3.
	
	\bibitem{Pan}
	C.~Pan and F.~R. Kschischang, ``Probabilistic 16-{QAM} shaping in {WDM}
	systems,'' \emph{IEEE/OSA J.\ Lightw.\ Technol.}, vol.~34, no.~18, pp.
	4285--4292, Sep. 2016.
	
	\bibitem{Fehenberger}
	T.~Fehenberger, A.~Alvarado, G.~B\"ocherer, and N.~Hanik, ``On probabilistic
	shaping of quadrature amplitude modulation for the nonlinear fiber channel,''
	\emph{IEEE/OSA J.\ Lightw.\ Technol.}, vol.~34, no.~21, pp. 5063--5073, Nov.
	2016.
	
	\bibitem{Buchali_2016}
	{F. Buchali}, {F. Steiner}, {G. B\"ocherer}, {L. Schmalen}, {P. Schulte}, and
	{W. Idler}, ``Rate adaptation and reach increase by probabilistically shaped
	64-{QAM}: An experimental demonstration,'' \emph{IEEE/OSA J.\ Lightw.\
		Technol.}, vol.~34, no.~7, pp. 1599--1609, Apr. 2016.
	
	\bibitem{Gha17}
	A.~Ghazisaeidi, I.~{Fernandez de Jauregui Ruiz}, R.~Rios-M\"{u}ller,
	L.~Schmalen, P.~Tran, P.~Brindel, A.~C. Meseguer, Q.~Hu, F.~Buchali,
	G.~Charlet, and J.~Renaudier, ``Advanced {C$+$L}-band transoceanic
	transmission systems based on probabilistically shaped {PDM}-64{QAM},''
	\emph{IEEE/OSA J.\ Lightw.\ Technol.}, vol.~35, no.~7, pp. 1291--1299, Apr
	2017.
	
	\bibitem{pillai_2014_jlt}
	B.~S.~G. Pillai, B.~Sedighi, K.~Guan, N.~P. Anthapadmanabhan, W.~Shieh, K.~J.
	Hinton, and R.~S. Tucker, ``End-to-end energy modeling and analysis of
	long-haul coherent transmission systems,'' \emph{IEEE/OSA J.\ Lightw.\
		Technol.}, vol.~32, no.~18, pp. 3093--3111, Sep. 2014.
	
	\bibitem{Hager15}
	C.~H\"{a}ger, A.~{Graell i Amat}, F.~Br\"{a}nnstr\"{o}m, A.~Alvarado, and
	E.~Agrell, ``Terminated and tailbiting spatially-coupled codes with optimized
	bit mappings for spectrally efficient fiber-optical systems,'' \emph{IEEE/OSA
		J.\ Lightw.\ Technol.}, vol.~33, no.~7, pp. 1275--1285, Apr. 2015.
	
	\bibitem{staircase_frank}
	B.~P. Smith, A.~Farhood, A.~Hunt, F.~R. Kschischang, and J.~Lodge, ``Staircase
	codes: {FEC} for 100 {G}b/s {OTN},'' \emph{IEEE/OSA J.\ Lightw.\ Technol.},
	vol.~30, no.~1, pp. 110--117, Jan. 2012.
	
	\bibitem{christian1}
	C.~H\"{a}ger, A.~{Graell i Amat}, H.~D. Pfister, A.~Alvarado,
	F.~Br\"{a}nnstr\"{o}m, and E.~Agrell, ``On parameter optimization for
	staircase codes,'' in \emph{Proc.\ Optical Fiber Commun. Conf. (OFC)}, Los
	Angeles, CA, 2015, pp. 1--3.
	
	\bibitem{She17}
	A.~Sheikh, A.~{Graell i Amat}, and M.~Karlsson, ``Nonbinary staircase codes for
	spectrally and energy efficient fiber-optic systems,'' in \emph{Proc.\
		Optical Fiber Commun. Conf. (OFC)}, Los Angeles, CA, 2017.
	
	\bibitem{h_braided}
	Y.~Y. Jian, H.~D. Pfister, K.~R. Narayanan, R.~Rao, and R.~Mazahreh,
	``Iterative hard-decision decoding of braided {BCH} codes for high-speed
	optical communication,'' in \emph{Proc.\ IEEE Global Telecommun.\ Conf.
		(GLOBECOM)}, Atlanta, GA, 2013, pp. 2376--2381.
	
	\bibitem{Hag17}
	C.~H\"ager, H.~D. Pfister, A.~{Graell i Amat}, and F.~Br\"annstr\"om, ``Density
	evolution for deterministic generalized product codes on the binary erasure
	channel at high rates,'' \emph{IEEE Trans. Inf. Theory}, vol.~63, no.~7, pp.
	4357--4378, Jul. 2017.
	
	\bibitem{hager_coded_mod}
	C.~H\"{a}ger, A.~{Graell i Amat}, H.~D. Pfister, and F.~Brännström, ``Density
	evolution for deterministic generalized product codes with higher-order
	modulation,'' in \emph{Proc. Int. Symp. Turbo Codes \& Iterative Inf.
		Processing (ISTC)}, Brest, 2016, pp. 236--240.
	
	\bibitem{bocherer2017}
	\BIBentryALTinterwordspacing
	G.~B\"{o}cherer, ``Achievable rates for probabilistic shaping.'' [Online].
	Available: \url{http://arxiv.org/abs/arXiv:1707.01134.}
	\BIBentrySTDinterwordspacing
	
	\bibitem{Pog12}
	P.~Poggiolini, ``The {GN} model of non-linear propagation in uncompensated
	coherent optical systems,'' \emph{J. Lightw. Technol.}, vol.~30, no.~24, pp.
	3857--3879, Dec. 2012.
	
	\bibitem{kaplan1993information}
	G.~Kaplan and S.~Shamai, ``Information rates and error exponents of compound
	channels with application to antipodal signaling in a fading environment,''
	\emph{AEU. Archiv f{\"u}r Elektronik und {\"U}bertragungstechnik}, 1993.
	
	\bibitem{mismatch_lapidoth}
	A.~Lapidoth, ``Mismatched decoding and the multiple-access channel,''
	\emph{IEEE Trans.\ Inf.\ Theory}, vol.~42, no.~5, pp. 1439--1452, Sep. 1996.
	
	\bibitem{Gra53}
	F.~Gray, ``Pulse code communication,'' US Patent 2\,632\,058, 1953.
	
	\bibitem{She17b}
	A.~Sheikh, A.~{Graell i Amat}, and G.~Liva, ``On achievable information rates
	for coherent fiber-optic systems with hard decision decoding,'' in
	\emph{Proc.\ European Conf. Optical Communications (ECOC)}, Gothenburg,
	Sweden, Sep. 2017.
	
	\bibitem{SheJLT}
	------, ``Achievable information rates for coded modulation with hard decision
	decoding for coherent fiber-optic systems,'' \emph{IEEE/OSA J.\ Lightw.\
		Technol.}, vol.~23, no.~35, pp. 5069--5078, Dec. 2017.
	
	\bibitem{kschischang1993optimal}
	F.~R. Kschischang and S.~Pasupathy, ``Optimal nonuniform signaling for gaussian
	channels,'' \emph{IEEE Trans.\ Inf.\ Theory}, vol.~39, no.~3, pp. 913--929,
	May 1993.
	
	\bibitem{Boc11}
	G.~B\"ocherer and R.~Mathar, ``Matching dyadic distributions to channels,'' in
	\emph{Proc.\ Data Compression Conf. (DCC)}, Snowbird, UT, Mar. 2011, pp.
	23--32.
	
	\bibitem{shape2}
	S.~Baur and G.~B\"ocherer, ``Arithmetic distribution matching,'' in \emph{Proc.
		Int. ITG Conf. Systems, Communications and Coding (SCC)}, Hamburg, 2015, pp.
	1--6.
	
	\bibitem{shape3}
	R.~A. Amjad and G.~B\"ocherer, ``Fixed-to-variable length distribution
	matching,'' in \emph{Proc.\ IEEE Int.\ Symp.\ Inf.\ Theory (ISIT)}, Istanbul,
	2013, pp. 1511--1515.
	
	\bibitem{shapingGeorg}
	P.~Schulte and G.~B\"ocherer, ``Constant composition distribution matching,''
	\emph{IEEE Trans.\ Inf.\ Theory}, vol.~62, no.~1, pp. 430--434, Jan. 2016.
	
	\bibitem{bocherer2017highspeed}
	\BIBentryALTinterwordspacing
	F.~S. G.~B\"{o}cherer and P.~Schulte, ``High throughput probabilistic shaping
	with product distribution matching.'' [Online]. Available:
	\url{http://arxiv.org/abs/arXiv:1702.07510.}
	\BIBentrySTDinterwordspacing
	
	\bibitem{bochererECOChighspeed}
	------, ``Fast probabilistic shaping implementation for long-haul fiber-optic
	communication systems,'' in \emph{Proc.\ European Conf. Optical
		Communications (ECOC)}, Gothenburg, Sweden, Sep. 2017.
	
\end{thebibliography}
